\documentclass[a4paper,fleqn,usenatbib]{mnras}

\usepackage[T1]{fontenc}
\usepackage{ae,aecompl}
\usepackage{booktabs,caption}
\usepackage[flushleft]{threeparttable}
\usepackage{graphicx}	% Including figure files
\usepackage{amsmath}	% Advanced maths commands
\usepackage{amssymb}	% Extra maths symbols
\usepackage{float}
\usepackage{lastpage}

\title[USNO-B1.0 selection of blue--red objects]{Application of USNO-B1.0 towards selecting objects with displaced blue and red components}

\author[J.S. Jayson]{Joel S. Jayson$^{1}$\thanks{E-mail: jsjayson@yahoo.com}
\\
$^{1}$P.O. Box 34, Brooklyn, N.Y 11235, USA}

\pubyear{2015}

\begin{document}
\label{firstpage}
\pagerange{\pageref{firstpage}--\pageref{lastpage}}
\maketitle

% Abstract of the paper
\begin{abstract}
We have conducted a feasibility study to determine the effectiveness of using USNO-B1.0 data to preferentially detect objects with displaced red and blue components. A procedure was developed to search catalogue entries for such objects, which include M dwarfs paired with white dwarfs or with earlier main-sequence stars, and galaxies with asymmetric colour distributions. Residual differences between red and blue and infrared and blue scanned emulsion images define vectors, which, when appropriately aligned and of sufficient length, signal potential candidates. Test sample sets were analysed to evaluate the effective discrimination of the technique. Over 91,000 USNO-B1.0 catalogue entries at points throughout the celestial sphere were then filtered for acceptable combinations of entry observations and magnitudes and the resulting total of about 17,000 entries was winnowed down to a little more than 200 objects of interest.  These were screened by visual examination of photo images to a final total of 146 candidates.  About one quarter of these candidates coincide with SDSS data. Those constituents fall into two groups, single and paired objects. SDSS identified several galaxies in the first group. Regarding the second group, at least half of its members were tentatively identified as main-sequence pairs, the greater portion being of widely separated spectral types. Two white dwarf--main-sequence pairs were also identified. Most importantly, the vectors formed from  USNO-B1.0 residuals were in alignment with corresponding SDSS pair position angles, thereby supporting this work`s central thesis.

 \end{abstract}


\section{Introduction}

With USNO-B1.0 the United States Naval Observatory (USNO) presented a catalogue of over 7400 scanned Schmidt plates exposed in surveys that spanned a period from 1949 to 2002 \citep{Monet2003}. The catalogue represents a culmination of the photographic era of observation. 
We develop a means of selecting objects with displaced red and blue components through analysis of USNO-B1.0 entry fields. (The catalogue is hereafter referred to in the text as USNO-B).
\citet*{Christy1983} first proposed a method of detecting binaries with different coloured components by employing a shift in position dependent upon the observation band. Others \citep{Sorokin1985,Wielen1996,Bailey1998} further investigated the phenomenon and  \citet{Pourbaix2004} were first to successfully detect a number of binary systems on the basis of such colour differentiation, deriving their candidates from Sloan Digital Sky Survey (SDSS) Data Release 2 entries.  

Entities anticipated to be singled out include M dwarfs (dM) paired with early main-sequence stars, white dwarfs (WDs) paired with main-sequence stars and  galaxies with asymmetric colour distribution such as those with extensive stellar nurseries. The preferential selection of main-sequence pairs with widely spaced spectral types lends itself to studies of low mass ratio binaries \citep{Fischer1992,Clarke2009,Forgan2011,Reggiani2013,Duchene2013}. By-products of the procedure include distant blue objects, red giant/carbon stars or galaxies in angular proximity to main-sequence stars. In concept the technique, augmented with x-ray survey data, could further detect young neutron star--main-sequence binaries. However, the paucity of such objects makes this latter possibility unlikely. 

Central to the technique, the pair of stars or galaxy must appear as a single object in all observations of the USNO-B entry. (Though the proximity of two stellar objects is more likely to indicate a binary association than a chance occurrence, other than when discussing known binaries, we  generally use the term `pair'`or pairing'  to describe two objects in angular proximity). Our initial focus is on three test groups. We then follow-up by extracting paired and galactic candidates from USNO-B entries taken at spot locations throughout the celestial sphere. 

The USNO-B catalogue has over one billion object entries in two epochs, with two colour magnitude readings for the first epoch and three colour readings for the second epoch. The deep photographic surveys captured objects as faint as magnitude 21. The resulting image sizes range from about 2.5 arcsec across for a magnitude \textit{V}=20 object to approximately 12 arcsec for magnitude 13 \citep{King1977}. Because bright objects were saturated, entities brighter than \textit{V}=13 were replaced in the USNO-B catalogue by Tycho-2 readings \citep{Hog2000}, where available. Those replacement entries are not of interest here. 

To determine which observations belonged to an entry, USNO first passed a search aperture of 3 arcsec through the assembled digital files. Two or more points located within the aperture were entered as a record. Identification of objects with large proper motions (pms) required opening the search aperture and entailed a more involved procedure. Entries vary from a minimum of two observations to a maximum of five. See \citet{Monet2003} for further details. Residual values in the mean observation epoch are listed for each observation, i.e., the arcsec deviation of both RA~($\alpha$) and Dec~($\delta$) from their mean values (the RA deviation,15$\Delta \alpha \cos \delta$, is the negative \textit{x} residual). We identify the sought after paired and galactic candidates through analysis of the residuals. The catalogue, when supplemented with the survey logs\footnote{Most survey logs are available at the USNO web site, \url{www.usno.navy.mil/USNO/astrometry/optical-IR-prod/icas}. A complete set of survey logs was  furnished courtesy of David G. Monet, USNO, Flagstaff Station}, furnishes enough data, including computed pm, to enable retrieval of the observation coordinates. The dispersion at the mean observation epoch is about 0.12 arcsec. For a discussion of absolute errors see \citet*{Roeser2010}.

We would not expect the residuals to vary much from 0.12 arcsec for a single star. Also, because of the rigid relationship between main-sequence stellar mass, radius and radiative intensity, when a main-sequence pairing appears on the emulsion as a single compact object we expect the centre of the object image to follow the more massive star regardless of exposure colour. Again, the residuals are anticipated to be of the order of the measurement errors. However, a significant proviso to this conclusion obtains when the pair is measured as a single object, but appears on the emulsion as an extended object, i.e., two abutting or overlapping entities.  In that circumstance, if the two stars are of widely separated spectral types the red--blue displacement will be detected. 

For a WD--main-sequence pairing the rigid main-sequence relationship no longer applies. A hot WD emits at a greater intensity, yet because of its small radius, it is not necessarily more luminous. The blue exposure is displaced from the red exposure and that shift will lead to larger residual values. A similar argument applies for a galaxy with a displacement of large active and quiescent regions. 

That is the effect, in brief, that is central to our procedure. In fact, there are other factors that can lead to large residuals. The USNO in their effort to capture many objects erred on the side of gathering together data points that did not belong to the same entity. Often the giveaway for such a hybrid will be large residuals. None the less, a distinction can usually be made between legitimate objects and these hybrids by several means, e.g., comparing magnitudes at a given colour from one epoch to the other, seeking anomalies in data plotted in the observation epochs, and by use of images and finder charts from the USNO website.

SDSS has charted about 35 per cent of the sky in five colours and with unprecedented photometric precision \citep{York2000,Abazajian2003,Ahn2012}.  Its primary purpose was collecting massive data on quasars and galaxies. Adjuncts to the programme have generated extensive WD catalogues \citep{Kleinman2013}. In addition to the colour induced displacement work of \citet{Pourbaix2004}, \citet{Smolcic2004} identified compact WD--main-sequence binaries by analysis of colour magnitudes across several bands . Over 2,500 such systems have been catalogued \citep{Rebassa2012,Rebassa2013,Lifang2014}.  

USNO-B represents an older technology. Its photometry is over an order of magnitude coarser than that of SDSS.  However, USNO-B is a comprehensive catalogue that encompasses the entire celestial sphere, which is a major reason for pursuing this feasibility study.

Section~\ref{sec:criteria for detection of objects of interest} establishes criteria to detect candidate entries. In Section~\ref{sec:Test Samples and Analysis of Outliers} groups of known single stars, dM binaries and WD--dM binaries are probed to test our approach and outlier points are analysed. The entries studied in Section~\ref{sec:Test Samples and Analysis of Outliers} each have five observations. Section~\ref{sec:Criteria for USNO-B entries with three or four observations} considers candidate identification from USNO-B listings with four observations and for one combination of three observations. The insights gained in Sections~\ref{sec:Test Samples and Analysis of Outliers} and \ref{sec:Criteria for USNO-B entries with three or four observations}  are put to use in Section~\ref{sec:Probing USNO-B entries for candidate pairings and galaxies} where USNO-B entries from various sectors of the sky are sifted for objects of interest. Section~\ref{sec:SDSS data} analyses the segment of the candidate population coincident with SDSS data.  That data provides star/galaxy identification and, through use of colour--colour plots, affords tentative typing of candidate pairings. We present our conclusions in Section~\ref{sec:Conclusions}.

\section{criteria for detection of objects of interest}
\label{sec:criteria for detection of objects of interest}
\subsection{The USNO-B data set and a measure of a blue-red shift}
\label{sec:The USNO-B data set and a measure of a blue-red shift}

\begin{table}
	\centering
	\caption{Surveys used for USNO-B1.0.  SERC-I* supplements POSS-IIN and is an extension of the SERC-I survey \citep{Monet2003}.}
	\label{tab:surveys}
	\begin{tabular}{llcr} % four columns, alignment for each
		\hline
		${No.}$ & Name & ${\Lambda}$ (nm) & Obs. dates\\
		\hline
		0 & POSS-IO & 350-500 & 1949-1965\\
		1 & POSS-IE & 620-670 & 1949-1965\\
		2 & POSS-IIJ & 385-540 & 1985-2000\\
		3 & POSS-IIF & 610-690 & 1985-1999\\
		4 & SERC-J & 385-540 & 1978-1990\\
		5 & ESO-R & 630-690 & 1974-1994\\
		6 & AAO-R & 590-690 & 1985-1998\\
		7 & POSS-IIN & 730-900 & 1989-2000\\
		8 & SERC-I & 715-900 & 1978-2002\\
		9 & SERC-I* & 715-900 & 1981-2002\\
		\hline
	\end{tabular}
\end{table}

\begin{figure*}
 \includegraphics[scale=.40]{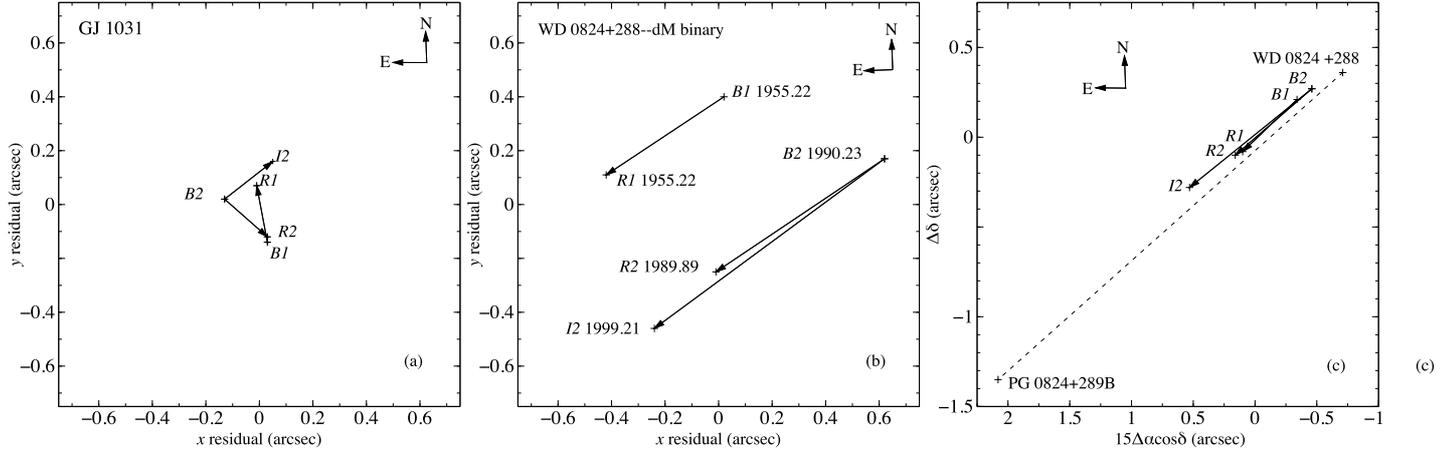}
 \caption{Residuals for a single star and for a WD--dM binary. The residuals are plotted in the mean epoch of observation in panes (a) and (b), 1972.9 and 1977.9, respectively.  They are plotted on the same scale for ready comparison. The observation dates for the WD--dM binary are noted in pane (b). These are put to use for the transformation indicated in pane (c). The origin in pane (c) is at the USNO-B1.0 1187-0163669 ICRS, epoch 2003.07 position, 08$^{h}$27$^{m}$05$\overset{^s}{.}$10+ 28$\degr$44$\overset{'}{}$02$\overset{''}{.}$4 as determined using GAVO corrections and a computed pm. See text for further discussion.}
 \label{fig:Fig1}
 \end{figure*}

Table~\ref{tab:surveys} summarizes the various surveys that contributed to USNO-B and is adapted from the USNO-B catalogue. A USNO-B entry contains 51 fields. Assuming that there is a full complement of five observations, about half of those fields are of immediate interest. Table~\ref{tab:fig1data} provides abbreviated entries for the two objects depicted in Fig.~\ref{fig:Fig1}. An alternate object designation (not provided in the USNO-B catalogue), and the object coordinates constitute the first three rows, followed by the mean observation epoch and the USNO computed pm components. All succeeding rows refer to individual observations. They include the survey number, the survey field, the date of the observation (not provided in the USNO-B catalogue, obtained from the survey log), the \textit{x} residual and the \textit{y} residual.
\begin{table}
	\centering
	\caption{Data for two objects plotted in Fig. 1.  USNO-B1.0 0611-0011923 is a single star. USNO-B1.0 1187-0163669 is a WD--dM binary whose components are WD 0824+288 and PG 0824+289B.  Coordinates are at equinox J2000, epoch J2000.}
	\label{tab:fig1data}
	\begin{tabular}{lll}
		\hline
		& \multicolumn{2}{c}{USNO-B1.0 identities}\\
		& 0611-0011923 & 1187-0163669\\
		\hline
		Alt. Designation & GJ 1031 & WD 0824+288 $\&$ dM\\ 
		 RA ($^h$ $^m$ $\overset{^s}{.}$) & 01 08 18.28 & 08 27 05.14\\
		Dec ($\degr$  $\overset{'}{}$  $\overset{''}{.}$ )& -28 48 20.09 & 28 44 02.65\\
		Mean epoch of obs. & 1972.9 & 1977.9\\
        		RA pm & 724~mas yr$^{-1}$ & 0~mas yr$^{-1}$\\
		Dec pm & -110~mas yr$^{-1}$ & 0~mas yr$^{-1}$\\
		\textit{B1} survey & 0 & 0\\
		\textit{B1} field & 884 & 312\\
		\textit{B1} obs. date & 1955.9 & 1955.2\\
		\textit{B1 x} residual & 0.03 arcsec & 0.02 arcsec\\
		\textit{B1 y} residual & -0.14 arcsec & 0.40 arcsec\\
		\textit{R1} survey & 1 & 1\\
		\textit{R1} field & 884 & 312\\
		\textit{R1} obs. date & 1955.9 & 1955.2\\
		\textit{R1 x} residual & -0.01 arcsec & -0.42 arcsec\\
		\textit{R1 y} residual & 0.07 arcsec & 0.11 arcsec\\
		\textit{B2} survey & 4 & 2\\
		\textit{B2} field & 412 & 431\\
		\textit{B2} obs. date & 1977.6 & 1990.2\\
		\textit{B2 x} residual & -0.13 arcsec & 0.62 arcsec\\
		\textit{B2 y} residual & 0.02 arcsec & 0.17 arcsec\\
		\textit{R2} survey & 6 & 3\\
		\textit{R2} field & 412 & 431\\
		\textit{R2} obs. date & 1994,6 & 1989.7\\
		\textit{R2 x} residual & 0.03 arcsec & -0.01 arcsec\\
		\textit{R2 y} residual & -0.12 arcsec & -0.25 arcsec\\
		\textit{I2} survey & 8 & 7\\
		\textit{I2} field & 412 & 431\\
		\textit{I2} obs. date & 1980.7 & 1999.2\\
		\textit{I2 x} residual & 0.05 arcsec & -0.24 arcsec\\
		\textit{I2 y} residual & 0.16 arcsec & -0.46 arcsec\\
		\hline
	\end{tabular}
\end{table}

A most useful measure gauges the relative displacement between the blue images and the corresponding red images and also, with regard to second epoch exposures, the relative displacement between the second epoch blue and infrared images. Concerning that relative displacement, consider surveys 0 and 1. Those are the blue and red surveys, respectively, of the first Palomar Observatory Sky Survey (POSS-I, see \citet{Minkowski1963}) . What is unique about  these readings is that the blue and red plates for each field were exposed on the same night. Thus, beyond the measurement error, the difference between the red and blue \textit{x} and \textit{y} readings gives a direct indication of the displacement between the two images in the observation epoch, as well as any other epoch that they both may be transformed to. For all other surveys, the difference in exposure dates between different colour plates within the same epoch vary from as little as a negligible amount to as much as 20 yr. Since the residuals are given in the mean epoch of the object observations, in theory the difference in exposure dates is transformed away. However, that transformation depends upon the USNO-B computed pm. Any error in computed pm will translate to a corresponding error in location, the error being equal to the product of the pm error times the difference in observation dates.  Large pm's play a role in the vicinity of the solar neighbourhood.  However, a good fraction of the USNO-B entries constitute distant objects with zero or close to zero pm.

 In Fig.~\ref{fig:Fig1} the residuals are plotted for two very different objects. In Fig.~\ref{fig:Fig1}a they are plotted for GJ 1031, a single star. Here \textit{R1} and \textit{B1} refer, respectively, to the epoch 1 red and blue exposure residual values and \textit{R2}, \textit{B2} and \textit{I2} are the corresponding epoch 2 red, blue and infrared residual values. Displacement vectors are drawn for \textit{\textbf{R1-B1}}, \textit{\textbf{R2-B2}} and \textit{\textbf{I2-B2}}, with the tail of the vector at the blue position. The data points are clustered within a small area and the vectors are directed in random directions. 
 
 In Fig.~\ref{fig:Fig1}b the residuals are plotted for a WD--dM binary. The WD component is WD 0824+288 and the dM component PG 0824+289B. In this instance the three vectors are aligned in nearly the same direction and the magnitudes of the vectors are significantly larger than those of Fig.~\ref{fig:Fig1}a.
 
Table~\ref{tab:fig1data} lists a USNO-B pm value of zero for the WD 0824+288, PG 0824+289B system. A glance at Fig.~\ref{fig:Fig1}b indicates that the pm value is in fact not zero. We surmise that the USNO-B algorithm may have set the pm to zero because of a contradiction in the RA terms. 

To demonstrate the relative position of the displacement vectors with respect to the two stars, we   compute an approximate corrected pm and transform all observations to ICRS, epoch 2003.07 to put them in the same reference frame and epoch as that of the stellar positions taken from SDSS data release nine.  For comparison with the SDSS data the  USNO-B coordinates require corrections to the ICRS frame  \citep{Roeser2010}, which can be found at the German Astrophysical Virtual Observatory (GAVO)\footnote{\url{http://dc.zah.uni-heidelberg.de/ppmxl/q/corr/info}}. Since the USNO-B pm is zero, the residuals represent the relative positions in the observation epochs as well as in the mean epoch. We compute an approximate pm by taking the differences of the second epoch blue and red observations with their corresponding observations in the first epoch, \textit{B2-B1} and \textit{R2-R1}, dividing by the differences in observation dates and taking a weighed average. Using the computed value of 16~mas yr$^{-1}$ at a position angle of 240\degr, the residuals are transformed and plotted in Fig.~\ref{fig:Fig1}c. The displacement vectors align in good agreement with the WD--dM position angle.  The slight offset between the displacement vectors and the binary reflects the approximate nature of the pm computation and the bias of the displacement vectors towards the WD signifies that that star is much brighter than its dM companion.

The WD 0824+288, PG 0824+289B system provides an introduction into what can be extracted from USNO-B. We next develop the specific criteria for gleaning candidate objects from the general population of USNO-B entries.

\subsection{A Point Count Approach for Evaluating Potential Candidates}
\label{sec:A Point Count Approach for Evaluating Potential Candidates}

 Though the displacement vectors in Figs.~\ref{fig:Fig1}b and \ref{fig:Fig1}c present a clear picture of the binary system, they embody a `noise' component as well as a `signal', Fig.~\ref{fig:Fig1}a providing some idea of that noise level. To get a more quantitative reading, the mean and sample deviation of the rms residual radius are computed for a sample of 40 single stars where, 
\begin{equation}
   r_{rms}=sqrt{((x_{1}^2+y_{1}^2+x_{2}^2+y_{2}^2+...x_{5}^2+y_{5}^2)/5)}.
   \end{equation}
  In this equation \textit{x}$_1$, \textit{x}$_2$...and \text{y}$_1$, \text{y}$_2$... represent the negative RA residual component (15$\Delta \alpha \cos \delta$) and the Dec residual component, respectively, of the observations that constitute an entry.  The sample stars all have five observations 
 
   The USNO-B identity, a standard identity, coordinates and the computed rms residual radius for each of these stars is provided in Table~\ref{tab:singlestar}. The last column of this table, listing point counts, is addressed in Section~\ref{sec:Test Samples and Analysis of Outliers}. To ensure that the stars in this group are indeed single, the sample was selected from the solar neighbourhood, no further than 20 pc, a region where extensive exploration extends some confidence. [One object, GJ 4360, is a barely resolved dM binary \citep{Montagnier2006}, which with an angular separation of $\sim$0.1 arcsec, for all practical purposes stands in as a single star in this exercise.] The selection was dictated in large part by brightness, since many stars at that close range were replaced in the USNO catalogue by Tycho-2 readings. dMs predominate along with a couple of WDs. Since only USNO-B entries with five observations were included that further limited the selection sample. The mean \textit{r}$_{rms}$ for the sample falls very much as expected, at 0.14~arcsec with a sample deviation of 0.05~arcsec.

Referring to Figs.~\ref{fig:Fig1}b and \ref{fig:Fig1}c, we readily spell out the requirements for an entry with five observations:
\begin{enumerate}
\item The magnitude of each displacement vector should be significantly larger than 0.15 arcsec, yet not so restricted as to to eliminate closely spaced pairs. We choose a minimum of 0.3 arcsec.
\item The angle between any two displacement vectors must fall below some maximum cutoff. Here we use a nominal noise level of 0.15 arcsec and for each displacement vector (\textit{\textbf{DV}}) allow an angular deviation of arctan(0.15/|\textit{\textbf{DV}}|). That expression is obtained by assuming a worst case of the noise component in quadrature with the signal component. By adding the maximum allowed deviation from alignment for each of two vectors we arrive at a comfortable margin of error where the allowed angular deviation between two displacement vectors must be less than arctan(0.15/|\textit{\textbf{DV1}}|)+arctan(0.15/|\textit{\textbf{DV2}}|). The larger the vectors, the tighter the angular deviation restriction. Alternatively, a small displacement vector magnitude leads to a loose restriction. We address that problem below.
\item As seen in Figs.~\ref{fig:Fig1}b and \ref{fig:Fig1}c, if the object is a legitimate candidate then, |\textit{\textbf{I2-B2}}|> |\textit{\textbf{R2-B2}}|.
\end{enumerate}

A point system with these requirements leads to a maximum of three points for satisfying minimum displacement vector magnitudes and a maximum of three points for satisfying the condition on restricted angular deviations.  We address condition (iii) as follows: |\textit{\textbf{I2-B2}}| >|\textit{\textbf{R2-B2}}| =1 point; 0.2 arcsec>|\textit{\textbf{R2-B2}}|-|\textit{\textbf{I2-B2}}|>0=0 points; |\textit{\textbf{R2-B2}}|-|\textit{\textbf{I2-B2}}|>0.2 arcsec= -1 point. Regarding small displacement vector magnitudes, if any |\textit{\textbf{DV}}| <0.15 arcsec, then for the two angles that that vector forms with the other two vectors, the angular deviation point count is zero. The displacement vectors are limited to residual differences between observations in the same epoch. See Section~\ref{sec:Mixed epoch vectors} for a short discussion of mixed epoch vectors. 

Before putting these requirements to trial with a population of unknowns, we evaluate them first on known quantities in the following section.

\section{Test Samples and Analysis of Outliers}
\label{sec:Test Samples and Analysis of Outliers}

We anticipate that our point system will select paired main-sequence stars with extended USNO-B images and of widely separated spectral types. WD--dM pairs are expected to be selected, even if their USNO-B images are compact. However, selection of dM binaries is not considered likely regardless of emulsion image extension. Three sample groups, each of 40 objects, are evaluated to test these expectations, single stars, dM binaries and WD--dM binaries. Each object comprises five observations. The single star group is the same as that represented in Table~\ref{tab:singlestar} and is included as a benchmark.  The binary sample group members are listed in Table~\ref{tab:binaries}.

\cite{Janson2012,Janson2014} have conducted a large-scale study of multiple dM systems using earth bound telescopes, but obtaining diffraction limited resolution through the use of speckle imaging. Table~\ref{tab:binaries} provides the USNO-B identity, a standard identity for one of the components, coordinates, the component separation and the computed point counts for 40 of these systems.  Systems were selected with separations greater than a few tenths of an arcsec and less than 7.5 arcsec. 

The sample of WD--dM binaries, found in Table~\ref{tab:binaries}, derived from several sources. \citet{Farihi2006,Farihi2010} conducted a study with the \textit{Hubble Space Telescope} in which over forty systems were resolved into two or more components. \citet{Hoard2007} established a table of WD--low mass star binaries acquired from a culmination of work entailing analysis of the \citet{McCook1999} catalogue using 2 Micron All Sky Survey (2MASS) photometry, \citet{Skrutskie2006} and follow up telescopic surveys. \citet{Silvestri2005} compiled a table of WD--dM wide binaries in the course of a study on the age of such systems. In all these instances we selected samples with separation distances from several tenths of an arcsec to 10 arcsec. Table~\ref{tab:binaries} indicates whether or not SDSS data is available for the WD--dM entries.

\begin{figure*}
 \includegraphics[scale=.50]{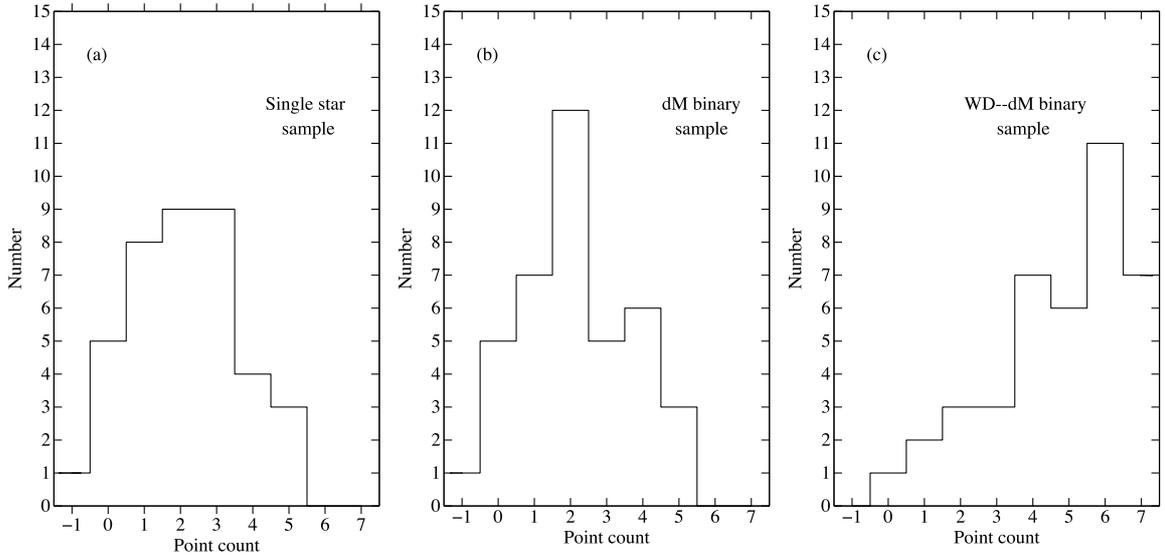}
 \caption{Histograms for single star, dM binary and WD--dM binary sample sets.  The WD--dM sample set shows a distinct bias towards higher point counts.}
 \label{fig:Fig2}
 \end{figure*}

Fig.~\ref{fig:Fig2} provides histograms for the three groups of samples, the single stars in Fig.~\ref{fig:Fig2}a, the binary M-dwarfs in Fig.~\ref{fig:Fig2}b and the WD--dM binaries in Fig.~\ref{fig:Fig2}c. There are no point counts of 6 or 7 for the first two groups, while the third, the group of WD--dM binaries, has a total of 18 objects at those levels and an additional 6 objects with a point count of 5. That is an encouraging result, but it leaves open the question of why the contrast is not even more extreme. The outliers are investigated in the following three subsections. These are the objects in the first two groups with point counts of 4 and 5, and the objects in the third group with point counts 5 and lower. 

\subsection{Single Star Outliers}
\label{sec:Single Star Outliers} 

In Fig.~\ref{fig:Fig2}a four stars exhibit a point count of 4 and three a point count of 5. Two of the stars with point count 5 and one of the stars with point count 4 exhibit diffraction spikes on the USNO-B images. \citet{Monet2003}, in explaining the reason for substitution of Tycho-2 entries for bright stars, state that the USNO precision measuring machine's measures are `usually confused by\ldots gross saturation, diffraction spikes and halos.' Eliminating from consideration USNO-B bright stars that weren't captured by Tycho-2 would partially resolve this problem. That option is adopted in Section~\ref{sec:Probing USNO-B entries for candidate pairings and galaxies}. Diffraction spikes from neighbouring stars remain an issue.

At the other extreme, LP 888-18, a star with point count 5,  has a \textit{B1} magnitude of 20.71 and a  \textit{B2} magnitude of 19.96. We attribute the large displacement vectors and high point count of this star to the less precise measurements at the higher magnitudes. An upper limit of 19.5 \textit{B2} magnitude will be introduced in section 5. Large spreads in magnitude, such as in the blue in this instance, are not uncommon in USNO-B. \citet{Monet2003} give a standard deviation of 0.25 mag. For eligibility into the sample groups objects were required to differ by less than one magnitude in the red and in the blue.

A large fraction of the stars in Fig.~\ref{fig:Fig2}a have point counts of 1, 2 or 3. These stars have short displacement vectors, but not necessarily shorter than the angular deviation cutoff of 0.15 arcsec. As a result moderate sized angular deviations are awarded points. Additionally, if the residuals are solely due to measurement errors we can expect that for half of these objects |\textit{\textbf{I2-B2}}|>|\textit{\textbf{R2-B2}}|, and hence another point will be awarded. It is therefore not surprising that a few objects with rms residual radii in the vicinity of 0.12 arcsec none the less register point counts of four. The remaining three objects fall into this category.

However, that is not the full extent of the problem. All three of these stars have an inclusive angle of less than 11$\degr$  between the \textit{\textbf{R2-B2}} and \textit{\textbf{I2-B2}} vectors. For that matter, six of the seven outlier stars fall into this category and of the forty stars in the sample, twelve have an inclusive angle of less than 15\degr. By comparison, for the \textit{\textbf{R1-B1}}and \textit{\textbf{I2-B2}} vectors, four of the forty stars satisfy that condition and for the \textit{\textbf{R1-B1}} and \textit{\textbf{R2-B2}} vectors two out of forty do.

\textit{\textbf{R2-B2}} and \textit{\textbf{I2-B2}}  are not totally independent vectors, both sharing the common \textit{B2} point. Given three observation points, and deviations due to measurement errors alone, assume one of the measurements more errant than the other two. When that point is \textit{B2}, we expect that the two vectors formed from \textit{B2} will be closely aligned. This explanation satisfies the statistics, although the sample size is limited.  This is a serious problem if the number of false candidates is to be minimized. It is especially so when, as will be discussed in Section~\ref{sec:Criteria for USNO-B entries with three or four observations}, searches are to be conducted using only the two epoch 2 vectors. We expect that the longer the displacement vectors, the less likely the effect.  Aside from requiring longer displacement vectors, a cap on the \textit{B2} magnitude, discussed above, will further alleviate the problem.

\subsection{dM binary outliers}
\label{sec:dM binary outliers}
In Fig.~\ref{fig:Fig2}b six objects have a point count 4 and three a point count 5. Two of the point count 4 objects display prominent diffraction spikes as does one with a point count of 5. Another point count 4 binary falls into the category of an object with small residuals resulting in random angular deviations that are awarded points.

The other five objects of interest have large point counts because the passage of a neighbouring star, not a member of the binary, affects the results. We take a close look at two of these objects.

\subsubsection{2MASS J05191382-0059423}

\begin{figure*}
 \includegraphics[scale=.50]{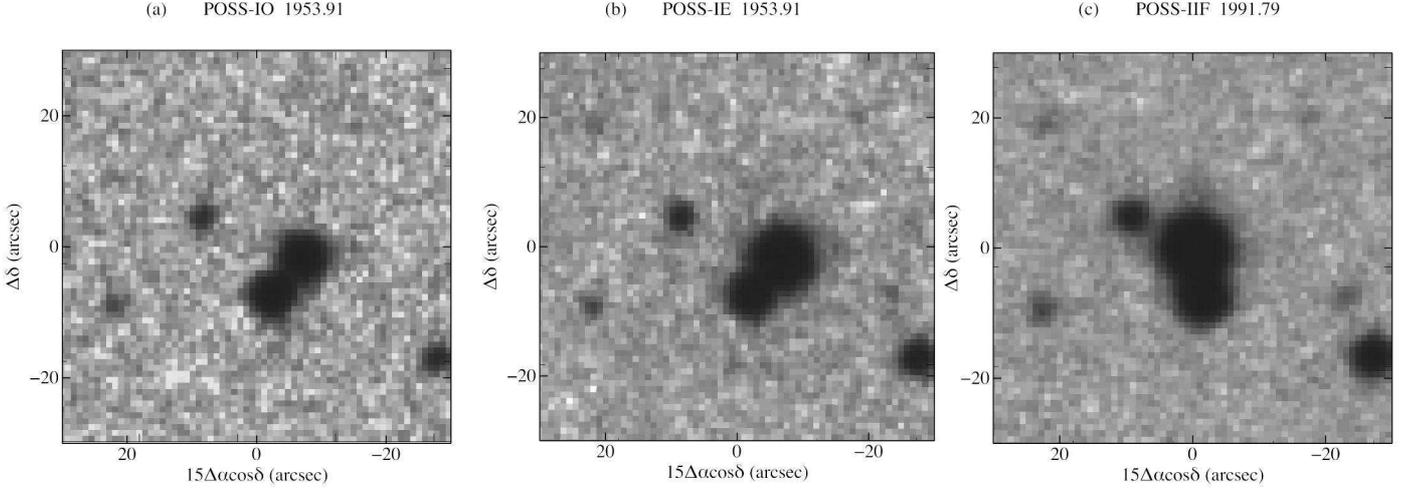}
 \caption{USNO-B1.0 images for 2MASS J05191382-0059423.  All images are shown in equinox J2000, epoch J2000.  The observation dates are indicated above each pane.  The centre coordinate for each is 05$^h$19$^m$14$\overset{^s}{.}$35 -00$\degr$59$\overset{'}{}$38$\overset{''}{.}$0.  See text for detailed description.}
 \label{fig:Fig3}
 \end{figure*}

Figure~\ref{fig:Fig3} depicts downloaded USNO images at the 2MASS J05191382-0059423 position. In the 0 and 1 survey images, shown respectively in Figs.~\ref{fig:Fig3}a and \ref{fig:Fig3}b, the object identified by USNO-B is in fact two objects. The segment to the southeast (north is up, east is to the left) exhibits little change from the blue image to red. The second segment increases its size significantly on the red exposure. It is that object that has the coordinates of 2MASS J05191382-0059423 and within that object lies the binary system, with a separation of 1.1 arcsec, that we originally set out to probe. Further, Fig~\ref{fig:Fig3}c displays the image from survey 3. The first two surveys were conducted in 1953.91. Survey 3 was conducted in 1991.79 and indicates either the apparent orbiting of the `blue' and `red' objects or one object passing the other. They have an angular separation of  roughly $\sim$7 arcsec. According to \citet{Janson2014} the distance of the `red' object is about 31 pc, which, if the objects are part of the same system, translates to a separation of $\sim$200 au.  The total mass of the two red companions is 0.65 M$_{\odot}$ \citep{Janson2012} and if the objects are gravitationally bound the `blue' object is likely a WD with mass of the order of M$_{\odot}$ and the system has a period on the order of 3000 yr. In a span of 38 yr the objects have changed their orientation by about 45 degrees. We are probably observing a fly-by. The `blue' object has SDSS and 2MASS identities, SDSS J051913.68-005949.9 and 2MASS J05191367-0059499.

\subsubsection{2MASS J05464932-0757427}
\begin{figure}
 \includegraphics[width=\columnwidth]{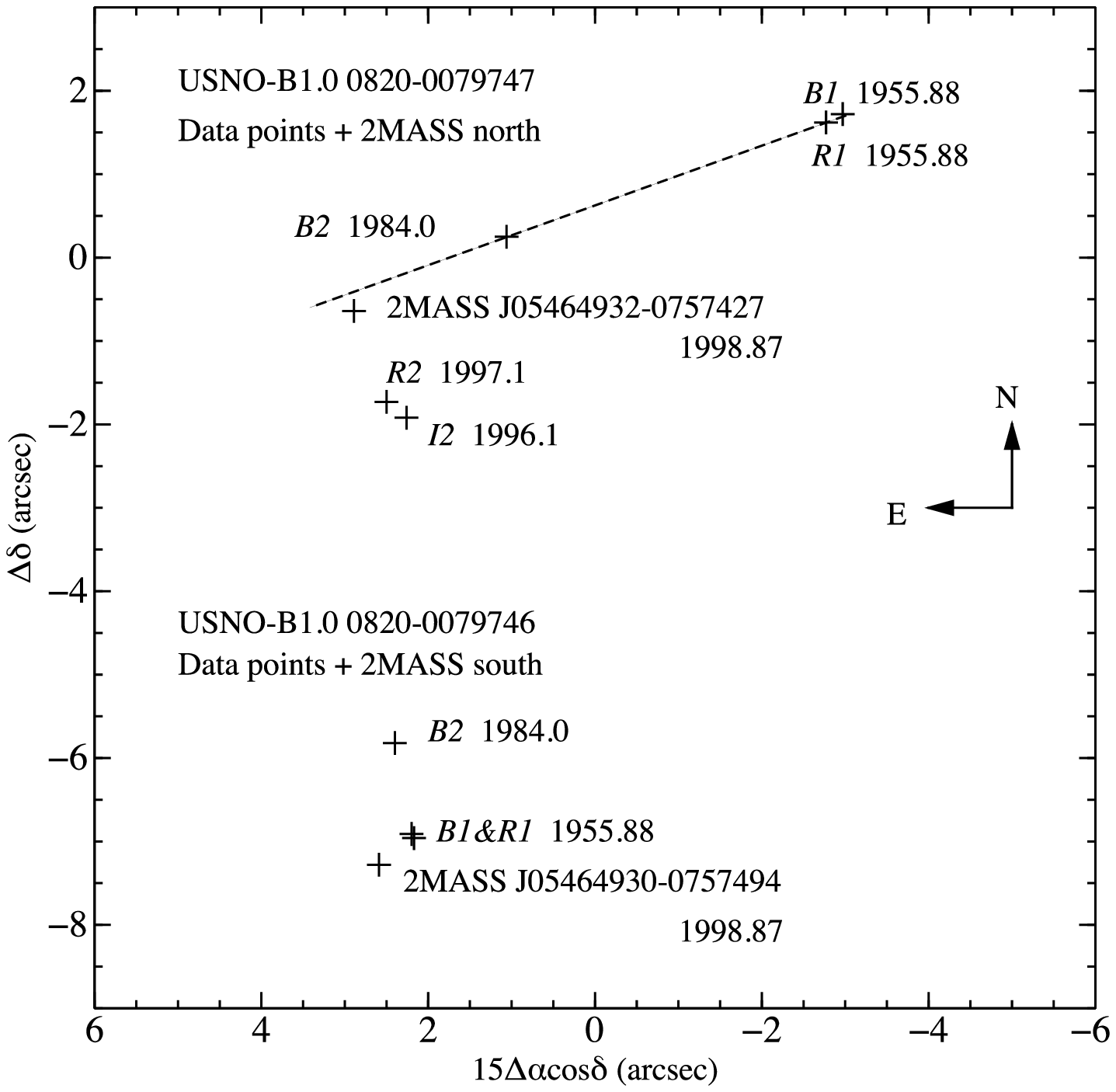}
 \caption{Plot in the observation epochs of data associated with 2MASS J05464932-0757427 and 2MASS J05464930-0757494. The origin is at the mean epoch for USNO-B1.0 0820-0079747, 1977.7, in the ICRS frame, with corrections obtained from GAVO \citep{Roeser2010}, 05$^h$46$^m$49$\overset{^s}{.}$13 -07$\degr$57$\overset{'}{}$42$\overset{''}{.}$1.  A dashed line is drawn through the \textit{B1, R1} and \textit{B2} observations. The line indicates the object direction of travel with increasing time and is consistent with the position of 2MASS J05464932-0757427 at the time of its observation in 1998.87.  The \textit{R2} and \textit{I2} observation positions are dislocated to the south as a result of inclusion in USNO-B1.0 0820-0079747 of the merged image of the northern and southern 2MASS stars at those times.}
 \label{fig:Fig4}
 \end{figure}

USNO-B images at the 2MASS J05464932-0757427 position indicate the blending of two star images with the passage of time.  A plot of the data in the observation epochs (Fig.~\ref{fig:Fig4}) delineates the situation. The epoch 2 red and infrared data points reflect the blended image while all other data points refer to the separate objects. USNO-B has two entries. The one associated with 2MASS J0546932-0757427 is USNO-B1.0 0820-0079747, which has five data points, and is the entry under analysis. The second entry, USNO-B1.0 0820-0079746 is associated with a southern object, 2MASS J05464930-0757494 and has three data points. The blending of the two images, with the concurrent southward shift of the \textit{R2} and \textit{I2} points results in two aligned and large epoch 2 displacement vectors, and a count of 4 points.

\subsection{WD--dM binary outliers}
\label{sec:WD--dM binary outliers}
For the WD--dM binary sample group the outliers are the entries with the low point counts as contrasted with the other two sample groups. Twenty-two  entries out of forty have point counts below 6. The low point count entries fall into a few major groups: large pair angular separation, small pair angular separation, large magnitude spread between the pair components, tight angular constraints between the displacement vectors and miscellaneous. A summary follows, again, a pair is referred to by just one component:
\begin{enumerate}
\item  Six entries with pair component angular separations ranging from 5 to 10 arcsec, display as separate entities on all observation images or as a mix, separate on at least one image and blended on at least one image. We require a blended object on all images for the displacement vectors to properly reflect the presence of red and blue components. Three entries in the sample group, WD 0325+263 and WD 1402+506, each with point count 6, and WD 2341-164 with point count 7, have separations of 6-6.3, 5 and 6.2-6.7 arcsec respectively, giving some indication of how large a separation can yield positive results. 
\item Three entries with pair component separations less than 0.7 arcsec failed to achieve 6 or 7 points. In all three instances the main problem lies with one or more short displacement vectors, which is not unexpected. Two entries in the sample group, WD 1443+337 and WD 1558+616, each with point counts of 6, have respective pair component separations of 0.68 and 0.72 arcsec, providing some measure of the minimal separation that can be detected.
\item For eight entries one stellar component dominates throughout the three colour spectrum, leading to a low point count. 
\item Two entries lost points because their large displacement vectors resulted in tight constraints on angular alignment. Thus, the position angles for the three WD 0949+451 vectors ranged from 102\degr to 124\degr, yet two points were deducted on this account. 
\item WD 1240+754 is highlighted here because of an artefact that appears, improbably, on the POSS-IE exposure as an extension to the binary system. WD 1619+525 is a multiple star system.  Two of the displacement vectors roughly align with the far companion. \textit{\textbf{R1-B1}} is over 120\degr out of alignment.  \cite{Hoard2007} made a tentative identification of WD 2318-137, point count 3, as a binary in their table 2 and this system would not have been considered for inclusion in the sample group except that the tentative assignment stemmed from observation of elongations in the POSS red images. We confirm moderate sized displacement vectors (0.25 to 0.48 arcsec). However, the vector position angles are unaligned.
\end{enumerate}

\section{Criteria for USNO-B entries with three or four observations}
\label{sec:Criteria for USNO-B entries with three or four observations}

\subsection{Two displacement vector criteria}
\label{sec:Two displacement vector}

Objects with five observations provide an opportunity for verification through several variables. Since there were no first epoch blue plate exposures taken south of -33$\degr$ declination we must also address four object detections. That is also true in the north if a greater portion of the catalogue is to be accessed.  Three object observations are also considered under one circumstance.

 No more than two displacement vectors can be formed from four observations. (An epoch 2 vector could be formed from \textit{\textbf{I2-R2}}, but that does not appear promising to us and will not be considered). Entries limited to N=3 can be assessed only in the instance when all three observations are from epoch 2, in which situation we can compare \textit{\textbf{R2-B2}} with \textit{\textbf{I2-B2}}. Beyond entries with five observations, the additional possibilities are: (\textit{B1,R1,B2,R2}), (\textit{B1,R1,B2,I2}), (\textit{B1,B2,R2,I2}), (\textit{R1,B2,R2,I2}) and (\textit{B2,R2,I2}).

With less than five observations there is little opportunity for assigning points. Two displacement vectors and their position angle difference yield three points. A fourth point is assigned for |\textit{\textbf{I2-B2}}|>|\textit{\textbf{R2-B2}}| for those entries with all three epoch 2 observations. That extra point in no way reflects the relative value of the pairing of two epoch 2 vectors versus an \textit{\textbf{R1-B1}} vector paired with an epoch 2 vector. The latter is a more reliable indicator as both vectors are completely independent. In all situations with fewer than five observations the minimum displacement vector magnitude is set at 0.45 arcsec to reduce the false candidate tally. Also, maximum point count, three or four, whichever applies is required. 

\subsection{Mixed epoch vectors}
\label{sec:Mixed epoch vectors}

We have chosen to take residual differences between emulsions from the same epoch. That was influenced to a great extent by difficulties with USNO-B pm values \citep{Gould2004,Munn2004}, as illustrated, for example, in Fig.~\ref{fig:Fig1}b. For distant objects, which constitute the great majority of USNO-B entries, the pm is either zero or close to it and there is no good reason why vectors cannot be formed from differences in residuals between images from both epochs. 

The introduction of vectors formed within and between epochs greatly increases the possibilities regarding combinations and numbers of vectors, but as long as each vector incorporates a blue observation the number of completely independent vectors will always be limited to two. Introducing mixed epoch vectors would allow analysis of entries that lack a \textit{B2} observation and would be particularly effective with four observation entries that include both blue observations. The option of mixed epoch vectors merits notice, but will not be pursued further in this work.

\section{Probing USNO-B entries for candidate pairings and galaxies}
\label{sec:Probing USNO-B entries for candidate pairings and galaxies}
The exercise in Section~\ref{sec:Test Samples and Analysis of Outliers} with sample groups of known quantities provided several insights germane to the more problematic task of probing the general population of USNO-B entries. The first step entails filtering out objects with unacceptable parameter values. Entries passing that gauntlet are then tested as described in the preceding sections and as amended below.

\subsection{The candidate selection sieve}
\label{sec:The candidate selection sieve }

The initial set of filters includes:
\begin{enumerate}
\item Eliminate all N=2 entries, all N=3 entries, except those with \textit{B2, R2} and \textit{I2} observations, and N=4 entries where \textit{B2} is the missing quantity.
\item Eliminate all entries where any magnitude is 21 or dimmer. This step filters against false readings caused by artefacts such as emulsion defects.
\item Eliminate all entries where $\Delta$ magnitude |\textit{B1-B2}|>0.9 and/or |\text{R1-R2}|>0.9. This requirement removes many of the hybrid entries and is slightly more exacting than the sample group requirement.
\item Include only entries in which 13<\textit{B2}<19.5 mag.  This requirement guards against diffraction spikes on the low end and against inaccurate readings on the high end.  It is a stringent requirement and on the high end, in and of itself, it eliminates a good fraction of all USNO-B entries.
\item Avoid densely populated areas, generally, but not necessarily, in the Galactic equatorial plane. Several of the outliers in Section~\ref{sec:dM binary outliers} experienced problems because of the passage of a neighbouring star.  The density in the vicinities of those outliers was 8 objects arcmin$^{-2}$ or greater. For this feasibility study we explore regions with densities <8 objects arcmin$^{-2}$. The average density of USNO-B entries over the celestial sphere is about 6.9 objects arcmin$\overset{^{-2}}{,}$ but the condition is restrictive due to the concentration about the Galactic plane.
\end{enumerate}

Additionally, rooted on the experience of the test sample groups, we tighten the requirements for entries with five observations. An entry must comply with each of the following conditions for selection:
\begin{enumerate}
\item All three displacement vectors must be greater than 0.35 arcsec.
\item Two of the three angle requirements between displacement vectors must be satisfied. That also puts a constraint on the third angle requirement.
\item |\textit{\textbf{I2-B2}}|-|\textit{\textbf{R2-B2}}|> -0.15 arcsec. This requirement combines and slightly modifies the three used with the sample groups.
\end{enumerate}

Based on these revised requirements, the displacement vector requirement, in particular, several of the WD--dM sample group entries that had point count 6 would not now be selected. This step was none the less taken in an attempt to minimize false positives. 

\begin{table*}
	\centering
	\caption{Summary of celestial sphere probes. See text for discussion of table content.}
	\label{tab:probes}
	\begin{tabular}{cccccccccccc} 
	\hline
	Galactic  &  & Acceptable  & Mag<21 &  & & N=4 & N=4 & N=3 & Density &\\
	coord. centre ($\degr$) & Entries & Obs. & diff.<0.9 & 13< \textit{B2}<19.5 & N=5 & \textit{R1+B1}  & \textit{R1} or 	\textit{B1} & epoch 2 & objs.~arcmin$^{-2}$ &\\
	\hline
Window 40~arcmin &  &  &  &  &  &  &  &  &  &\\ 
per side &  &  &  &  &  &  &  &  &  &\\ 
 &  &  &  &  &  &  &  &  &  &\\ 
0.0 +90.0 & 2335 & 887 & 515 & 174 & 2 & 0 & 1 & 0 & 1.5 &\\ 
0.0 +60.0 & 5073 & 2275 & 1250 & 510 & 2 & 0 & 1 & 2 & 3.2 &\\ 
120.0 +60.0 & 7669 & 2205 & 901 & 372 & 0 & 0 & 1 & 2 & 4.8 &\\ 
240.0 +60.0 & 2733 & 1256 & 1197 & 423 & 3 & 1 & 3 & 2 & 1.7 &\\ 
210.0 +45.0 & 2507 & 1145 & 872 & 405 & 0 & 0 & 1 & 0 & 1.6 &\\
60.0 +30.0 & 6962 & 3775 & 2679 & 1579 & 8 & 0 & 0 & 0 & 4.4 &\\ 
180.0 +30.0 & 5177 & 2493 & 1444 & 687 & 1 & 0 & 1 & 4 & 3.2 &\\ 
210.0 +30.0 & 4246 & 2304 & 1831 & 740 & 5 & 1 & 1 & 0 & 2.7 &\\
300.0 +30.0 & 9372 & 4431 & 3072 & 1583 & 1 & 0 & 7 & 1 & 5.9 &\\ 
0.0 -90.0 & 3486 & 1365 & 678 & 331 & 0 & 0 & 1 & 5 & 2.2 &\\ 
0.0 -60.0 & 3759 & 2131 & 2044 & 1068 & 0 & 0 & 18 & 1 & 2.4 &\\ 
120.0 -60.0 & 3659 & 1407 & 913 & 337 & 1 & 0 & 0 & 1 & 2.3 &\\ 
240.0 -60.0 & 3884 & 1592 & 983 & 524 & 0 & 0 & 8 & 0 & 2.4 &\\ 
60.0 -30.0 & 6843 & 4280 & 3118 & 1790 & 6 & 1 & 8 & 1 & 4.3 &\\ 
180.0 -30.0 & 3017 & 1533 & 1380 & 830 & 1 & 2 & 3 & 0 & 1.9 &\\ 
300.0 -30.0 & 9204 & 5076 & 3501 & 2075 & 0 & 0 & 56 & 1 & 5.8 &\\ 
 &  &  &  &  &  &  &  &  &  &\\ 
Window 20~arcmin &  &  &  &  &  &  &  &  &  &\\ 
per side &  &  &  &  &  &  &  &  &  &\\ 
 &  &  &  &  &  &  &  &  &  &\\  180.0 +0.0 & 2867 & 2059 & 1931 & 1053 & 11 & 2 & 2 & 0 & 7.2 &\\ 
90.0-15.0 & 2803 & 2047 & 1841 & 990 & 4 & 2 & 0 & 1 & 7.0 &\\ 
270+15 & 2600 & 1598 & 1391 & 920 & 0 & 0 & 9 & 2 & 6.5 & Total\\
15.0+15.0 & 3137 & 1985 & 1645 & 745 & 11 & 3 & 1 & 0 & 7.8 & objects of\\
 &  &  &  &  &  &  &  &  &  & interest\\
Totals & 91333 & 45844 & 33186 & 17136 & 56 & 12 & 122 & 23 &  & 213	\\	
\hline
	\end{tabular}
\end{table*}

\subsection{Candidates}
\label{sec:Candidates}
Table~\ref{tab:probes} summarizes which areas of the celestial sphere have been sampled. The first column provides the central Galactic coordinates and the search area about that centre. The next column lists the number of USNO-B entries found in the sample area as acquired from VizieR.  The following three columns provide the number of remaining entries after the filter at the column head has been activated, the value inclusive of all preceding filters. The next four columns indicate the number of objects of interest for objects with five observations, four observations and a displacement vector in each of the epochs, four observations and epoch 2 displacement vectors, and three observations, both vectors in epoch 2.  The final column gives the density of the initial entries~arcmin$^{-2}$. The row at the bottom of the table sums each of the columns. The total number of objects of interest is 213. This number represents unscreened results. That is out of an initial 91,133 entries.  The number of entries probed after filtering for acceptable observation detections and magnitudes is 17,136.

USNO-B images have been examined for all objects of interest. A total of 67 entries were removed from further consideration, overwhelmingly due to two problems: about two-thirds because of diffraction spikes from nearby bright objects and about one-third because of inconsistent object blending or separation from one emulsion to another. The removals were far from uniformly distributed. Of the 56 objects of interest with five observations  4 were removed. On the other hand, of the 122 items with four observations and two epoch 2 displacement vectors, 37 were removed. Further, of the 23 items with three observations and both displacement vectors in epoch 2, all 23 were withdrawn. Over half of those were phantom objects created out of the glare of the diffraction spikes. The diffraction spikes are consistently more prominent in the epoch 2 exposures than in those of epoch 1. Finally, of the 12 items with four observations and a displacement vector in each epoch, 3 were removed. These results highlight the robust nature of the entries with five observations.  As for the relatively large number of items with four observations and two vectors in epoch 2, that can partly be attributed to the absence of any \textit{B1} observations south of -33$\degr$ declination. Table~\ref{tab:list} provides a list of the 146 candidates remaining after screening. The first column furnishes the USNO-B identity, the next column the equatorial coordinates and the third column the number of observations.  Succeeding columns list the displacement vector position angles and vector magnitudes.  The final column notes USNO-B image characteristics and indicates availability of SDSS data.

\section{SDSS data}
\label{sec:SDSS data}
\subsection{WD--dM binary test set}
\label{sec:Binary test set}
Before considering SDSS data coincident with candidates in Table~\ref{tab:list}, we first return to the test set of WD--dM binaries, for which SDSS data is found for 30 of the 40 set members. Six of these members are flagged SATURATED, and are removed from consideration (see \citet{Stoughton2002} for a discussion of SDSS flagging).  An additional six entries are flagged BLENDED, no deblend. That comes as no surprise, as blended objects are central to this study.  However, interpreting the photometry of such objects is problematic. The SDSS telescopes have resolutions of about 1.5 arcsec, dependent upon seeing conditions.  The pairs, flagged BLENDED, no deblend, have separations within the range of 0.655 to 2.0 arcsec. For one of these binaries, WD 0956+045, only the dM is flagged. (Here, we again identify a binary by just one of its components, the WD).  

\begin{figure*}
 \includegraphics[scale=1.0]{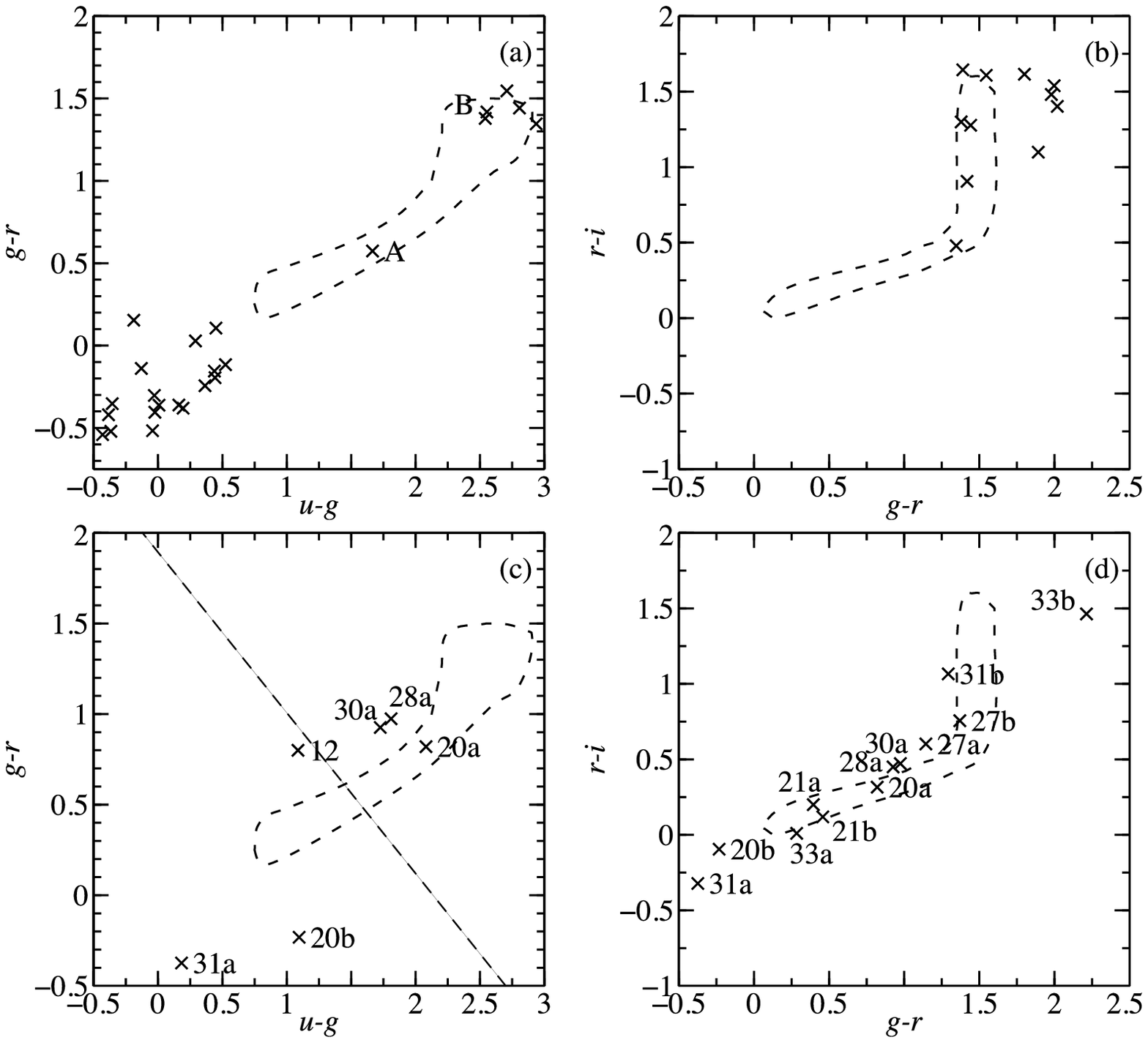}
 \caption{Colour--colour plots of WD--dM sample set pairs and of USNO-B1.0 candidate objects employing SDSS data. Fig.~\ref{fig:Fig5}a depicts WDs (lower left) from the WD--dM sample set and a few of their dM companions (upper right). WD 0725+827 (`A') and its dM companion, LP 5-74 (`B') are the only two stars identified on the plot. Most of the dM companions are plotted in Fig.~\ref{fig:Fig5}b. See Section~\ref{sec:Candidates coincident with SDSS data} for a discussion of objects plotted in Figs.~\ref{fig:Fig5}c and \ref{fig:Fig5}d.}
 \label{fig:Fig5}
 \end{figure*}
 
The colour--colour plots of Figs.~\ref{fig:Fig5}a and b display the remaining eighteen pairs and the unblended WD. Fig.~\ref{fig:Fig5}a plots \textit{g-r} versus \textit{u-g} and Fig.~\ref{fig:Fig5}b plots \textit{r-i} versus \textit{g-r}, where \textit{u, g, r} and \textit{i} are the ultraviolet, green, red  and infrared SDSS bands. (\textit{z} band data is not put to use here). The approximate locus for main-sequence stars is outlined on all diagrams \citep{Richards2002,Smolcic2004,Covey2007}. The WDs, with one exception, fall in the lower lefthand side of Fig.~\ref{fig:Fig5}a and the dM in the upper righthand side, both as expected from a wide body of work \citep{Lenz1998,Finlator2000,Smolcic2004,Covey2007,Rebassa2012,Rebassa2013}. The exception is the DC white dwarf, WD 0725+827 \citep{Silvestri2005}.  Cool WDs appear bluer than main-sequence stars at the same temperature \citep{Silvestri2002} and from its position on the colour--colour plot the WD 0725+827 temperature is less than 5500\degr C. Six of the binaries, WD 0257-005, WD 1106+316, WD 1157+129, WD1236-004, WD1558+616 and WD 1833+644, have been identified as a single star by SDSS, and lie within the WD region. Five of these binaries have separations ranging from 0.478 to 0.72 arcsec, which in this case was close enough to elude detection of blending. The sixth binary, WD 1833+644 is a multiple with separations of 0.079 and 1.82 arcsec. The nearer companion is 4 magnitudes brighter than the farther one \citep{Farihi2010}.

Only a few dMs are plotted on Fig.~\ref{fig:Fig5}a. The others exceed 20.5~mag in the \textit{u} band. At less than 20.5~mag the accuracy of the \textit{u} band measurements is better than 0.1~mag \citep{Smolcic2004}. [At 22 mag, the limiting signal-to-noise level is 5:1 \citep{Ivezic2000}, good enough for detection, but not for accurate photometry]. Eleven dMs are plotted in Fig~\ref{fig:Fig5}b. (One dM is too faint in the \textit{g} band to be plotted). Note that only WD 0725+827 is identified on Fig.~\ref{fig:Fig5}a.  No individual stars are identified on Fig.~\ref{fig:Fig5}b.

\subsection{Candidates coincident with SDSS data}
\label{sec:Candidates coincident with SDSS data}
SDSS observations coincide with 37 of the candidate entries.  Two of these observations have a SATURATION flag and one an EDGE flag.  The remaining entries are listed in Table~\ref{tab:SDSS}, representing a little less than a quarter of the total number of candidates.  Table~\ref{tab:SDSS} is divided into two parts, the first for single objects in which one SDSS item coincides with the USNO-B entry and the second part for paired objects in which SDSS has resolved or deblended two entities.  The first column of Table~\ref{tab:SDSS} numbers the entry for easy identification on Fig.~\ref{fig:Fig5}. That is succeeded by the USNO-B and SDSS identifications and by a column designating a  galaxy as indicated by SDSS. The following columns list the values of \textit{u}, \textit{u-g}, \textit{g-r}, and \textit{r-i}. For those  objects designated as galaxies by SDSS the values of \textit{u-g} and \textit{g-r}, corrected for reddening, as determined by \citet{Schlafly2011}, are parenthetically indicated adjacent to the uncorrected values. Their work has used SDSS data to improve upon the maps previously developed by \citet*{Schlegel1998}. Corrections have not been applied to  \textit{r-i} or to stellar objects.  Judging from the galactic results at the shorter wavelengths, such corrections for colour differences would not exceed 0.075~mag and their absence does not affect our conclusions. A second row in the paired object section records the same information for the second SDSS object.  The next two columns provide computed separation and position angles for the paired SDSS entries and the final column assesses the apparent nature of the object as determined from the SDSS designation and from the  colour--colour diagrams of Fig.~\ref{fig:Fig5}.   

\begin{table*}
\begin{threeparttable}
	\centering
	\caption{USNO-B1.0 entries coincident with SDSS data.  Tentative identifications are indexed as follows: (1) galaxy; (2) resolved WD--dM; (3) compact WD--dM paired with main-sequence star; (4) galaxy paired with faint object; (5) main-sequence star paired with apparent red giant; (6) galaxy BLENDED, no deblend; (7) galaxy paired with main-sequence star; (8) two main-sequence stars paired; (9) main-sequence star.  See text and Fig.~\ref{fig:Fig5} for a more detailed description.}

	\label{tab:SDSS}
	\begin{tabular}{cccccccccccc} 
	\hline
	
	 &  &  &  & \textit{u} & \textit{u-g} & \textit{g-r} & \textit{r-i} & sep. & PA & Tentative\\
$\#$ & USNO-B1.0 & SDSS & s/g & (mag) &(mag)  & (mag) & (mag) & (arcsec) & (\degr) &Identity\\
	\hline
1 & 0924-0009909 & J004422.53+022948.9 & galaxy & 20.317 & 0.775 (0.758) & 0.699 (0.682) & 0.324 & & &(1)\\
2 & 0970-0663419 & J212814.67+070050.4 & galaxy & 21.439 & 1.780 (1.717) & 1.314 (1.245) & 0.512 & & & (1)\\
3 & 0971-0681235 & J212726.08+071006.0 & galaxy & 20.604 & 1.042 (0.980) & 0.703 (0.635) & 0.357 & & &(1)\\
4 & 0972-0698506 & J212829.26+071614.2 &  & 20.077 & 1.695 & 0.671 & 0.217 & & & (9)\\
5 & 1006-0190377 & J110311.18+103932.2 & galaxy/BLD &  &  &  &  &  &  & (6)\\
6 & 1006-0190380 & J110313.16+103753.1 &  & 25.035 & 3.480 & 1.266 & 0.658  & & & \\
7 & 1009-0190583 & J110305.90+105657.0 &  & 23.100 & 1.810 & 1.529 & 0.789 & & & \\
8 & 1010-0189647 & J110317.05+110218.9 & galaxy & 22.256 & 0.991 (0.970) & 1.576 (1.553) & 0.748 & & &(1)\\
9 & 1010-0189899 & J110457.60+110150.0 &  & 18.927 & 1.219 & 0.505 & 0.209 & & & (9)\\
10 & 1170-0240143 & J125034.32+270059.1 &  & 20.722 & 1.039 & 0.574 & 0.217 & & & (9)\\
 &  &  &  &  &  &  &  &  &  &\\ 
11 & 0966-0584913 & J212750.95+063951.7 &  & 21.409 & 0.866 & 0.668 & 0.136 & 4.6&129 & (8)\\
 &  & J212751.19+063948.8 &  & 19.535 & 1.491 & 0.551 & 0.187 & & & \\
12 & 0969-0633655 & J212736.77+065924.5 & galaxy & 21.333 & 1.149 (1.087) & 0.867 (0.800) & 0.401 & & & (4)\\
 &  & J212736.59+065928.2 &  & 24.484 & 0.911 & 0.491 & -0.582 & & &\\ 
13 & 0969-0633980 & J212826.21+065837.3 &  & 19.344 & 1.552 & 0.581 & 0.247 & 3.0&90 & (8)\\
 &  & J212826.41+065837.3 &  & 21.043 & 2.617 & 1.383 & 0.645 & & &\\ 
14 & 0969-0634114 & J212851.73+065748.3 & galaxy & 21.260 & 1.505 (1.439) & 0.288 (0.215) & 0.122 &2.1& 203 & (7)\\
 &  & J212851.70+065746.3 &  & 22.275 & 1.947 & 1.050 & 0.324 & & &\\ 
 15  & 0971-0681479 & J212802.37+071015.8 &  & 23.098 & 3.226 & 1.341 & 0.578 & 4.7&280 &\\
 &  & J212802.68+071016.6 & star/BLD & 21.777$^a$ & -0.030$^a$ & 0.089$^a$ & -0.386$^a$ & & &\\ 
16 & 0971-0681705 & J212836.71+070849.8 & galaxy & 22.173 & 1.888 (1.822) & 1.449 (1.377) & 0.504 & 4.5&13 & (7)\\
 &  & J212836.64+070845.4 &  & 20.234 & 1.237 & 0.490 & 0.149 & & &\\ 
17 & 0984-0266003 & J143154.01+082559.3 &  & 19.455 & 2.464 & 1.080 & 0.401 & 5.4&233 &(7)\\
 &  & J143154.30+082602.6 & galaxy & 21.153 & 0.939 (0.918) & 0.229 (0.206) & 0.151 & & &\\ 
18 & 1009-0190822 & J110435.73+105746.1 & galaxy/BLD &  &  &  & & &  & (4), (6)\\
 &  & J110435.40+105740.7 &  & 23.214 & 1.325 & 1.192 & 0.512 & & &\\ 
 19&1053-0170335&J083536.69+151945.9 & &19.420 & 2.520 & 1.339 & 0.616  & 4.7&143 & (8)\\
&&J083536.49+151949.7&&17.743&1.303 & 0.483 & 0.163 & & &\\
20&1053-0170642&J083647.69+152210.7&&19.089&2.081&0.820&0.316&3.7&262& (2)\\
&&J083647.94+152211.2&&20.923&1.096&-0.232&-0.095&&&\\
21&1054-0169679&J083539.07+152536.8&&21.182&0.982&0.395&0.200&6.5&136& (8)\\
&&J083539.38+152532.1&&16.467&1.289&0.458&0.119&&&\\
22&1055-0172337&J083722.27+153038.0&&18.788&1.519&0.559&0.188&5.3&154& (8)\\
&&J083722.43+153033.2&&21.151&2.397&1.317&1.206&&&\\
23&1058-0172600&J083625.35+155111.5&&17.996&1.389&0.494&0.157&5.6&208& (8)\\
&&J083625.17+155106.6&&24.149&2.859&1.526&1.365&&&\\
24&1058-0172658&J083650.62+154859.0&&16.809&1.054&0.160&0.015&5.1&246& (8) \\
&&J083650.30+154856.9&&23.580&3.175&1.570&0.934&&&\\
25&1108-0181990&J093344.32+205300.1&&22.479&2.700&1.436&0.904&4.9&268& (7)\\
&&J093344.67+205300.3&galaxy&21.659&1.084 (1.056)&0.744 (0.713)&0.311&&&\\
26 & 1169-0233621 & J125133.53+265508.6 &  & 19.217 & 0.991  & 0.292 &  0.129 &4.4& 245&(8)\\
 &  & J125133.23+265506.7 &  & 25.542 & 2.661  & 1.546 &  1.659 &&&\\ 
27 & 1189-0098459 & J054534.11+285958.2 &  & 21.623 & 1.894 & 1.144 & 0.602 & 3.5&262&(8)\\
 &  & J054533.85+285957.7 &  & 24.901 & 3.602 & 1.370 & 0.756 &&&\\ 
28 & 1190-0098093 & J054521.60+290533.1 &  & 19.489 & 1.810 & 0.974 & 0.472 &7.75&79& (3)\\
 &  & J054521.02+290531.6 &  & 16.586 & 1.356 & 0.634 & 0.181 &&&\\ 
29 & 1190-0098147 & J054526.89+290542.8 &  & 19.154 & 1.991 & 0.753 & 0.383 &4.1 &274&(8)\\
 &  & J054526.58+290543.1 &  & 23.237 & 3.727 & 1.758 & 0.889 &&&\\ 
30 & 1190-0098184 & J054530.59+290507.8 &  & 19.533 & 1.726 & 0.926 & 0.450&4.2 &169& (3)\\
 &  & J054530.65+290503.7 &  & 22.668 & 2.506 & 1.561 & 1.239 &&&\\ 
31& 1251-0258748 & J173339.42+351123.4 &  & 19.092 & 0.183 & -0.374 & -0.322 &2.3&141& (2)\\
 &  & J173339.54+351121.6 &  & 25.321 & 4.769 & 1.293 & 1.066 &&&\\ 
32 & 1252-0258837 & J173435.38+351725.6 &  & 20.836 & 1.567 & 0.545 & 0.265 &7.2&290& (6), (7)\\
 &  & J173435.93+351723.1 & galaxy/BLD &  & &  &  &  &  &\\ 
33 & 1253-0259828 & J173420.48+351941.1 &  & 19.498 & 1.023 & 0.284 & 0.010 &1.8&251& (5)\\
 &  & J173420.34+351940.5 &  & 24.540 & 3.055 & 2.213 & 1.463 &&&\\ 
34 & 1254-0260615 & J173414.57+352745.0 &  & 19.848 & 1.248 & 0.471 & 0.158 &2.7&281& (8)\\
 &  & J173414.35+352745.5 &  & 25.108 & 4.575 & 1.381 & 0.690 &&&\\ 

\hline
	\end{tabular}
	\begin{tablenotes}
	\small
	\item \textit{Note} $^a$ Although the second component of object \#15 is flagged BLENDED, no deblend, the photometry is retained as discussed in the text.	
	\end{tablenotes}
	\end{threeparttable}\end{table*}

\subsubsection{Single objects}

The single objects include the first ten entries of Table~\ref{tab:SDSS} and additionally two entries from the second portion of the table, \#'s 12 \& 18, whose companions are too faint in all bands to have influenced USNO-B selection.  Both of those entries are identified as galaxies, for a total of seven of the twelve single objects designated as galaxies by SDSS . Two of those objects are flagged BLENDED, no deblend.  Regarding galaxy selection, our methodology works with distant systems. Galaxies that subtend tens of arc-seconds and more are either perceived by USNO-B as extremely bright objects (USNO-B magnitude evaluation depends upon image diameter) or are often apprehended as a series of smaller objects.  In the first instance bright magnitude filtering eliminates the entry and in the second, should the object make its way through filtering and analysis, it then should be eliminated on visual inspection.  The two galaxies flagged BLENDED, no deblend, are not nearby, their images span only a few arc-seconds, but they none the less fall into the latter category and eluded being discarded in the initial visual inspection. The other five galaxies appear as small images.  Entry \#12 is plotted on Fig.~\ref{fig:Fig5}c and falls close to the separator line between red and blue galaxies (dashed line, from \citet{Strateva2001}). That is suggestive of a bicoloured galaxy, but the significance of the selection of this galaxy, as well as the other four, requires observational input. \citet{Strateva2001} selected a population of galaxies with a cutoff of \textit{g}<22~mag. Entry \#12 satisfies that condition, but the \textit{u} band accuracy is probably no better than 0.2~mag.

Of the remaining five single object entries two, \#'s 6 and 7, have faint \textit{u} and \textit{g} magnitudes and each has four USNO-B observations, i.e., two displacement vectors, both dependent upon \textit{B2}.  Although USNO-B \textit{B2} is less than 19.5~mag for both, that is at considerable variance with the SDSS data and no attempt is made to further identify these two entries. Object \#'s 4, 9 and 10 fall within the main-sequence locus. The significance of their selection, again, requires more specific observational data.

\subsubsection{Paired objects}

The salient point for the paired object portion of Table~\ref{tab:SDSS} revolves about a comparison between the position angle of the two SDSS components and the average position angle of the corresponding USNO-B displacement vectors (see Table~\ref{tab:list} for displacement vector data). Out of a total of 22 items we find the following: 
\begin{description}
\item Fifteen entries agree within 9\degr, ten of those within 5\degr.
\item Six entries agree within a range of 11\degr-- 24\degr.
\end{description}

The remaining  entry, \#17, USNO-B 0984-0266003, deviates from alignment by 31\degr.  The separations for the 22 entries range from 1.8--7.75 arcsec, in accord with expectations.

Turning to the tentative identification of the entries, colour difference analysis, i.e., \textit{g-r} versus \textit{u-g} and \textit{r-i} versus \textit{g-r}, plus SDSS star/galaxy discrimination leads to the following conclusions: 
\begin{enumerate}
\item Eleven entries are main-sequence pairs, eight of them an dM paired with an earlier star.  All eleven pairs exhibit extended USNO-B images. Both components of two of the exceptions, \#'s 21 \& 27, USNO-B 1054-0169679 and USNO-B 1189-0098459, are plotted on Fig.~\ref{fig:Fig5}d. The close delta spectral spacings is of interest here. The spatial separations of the components, from Table~\ref{tab:SDSS}, are 6.5 and 3.5 arcsec, respectively.  To limit congestion, the third exception, \# 11, USNO-B 0966-0584913, was not plotted, but it is similarly closely spaced spectrally. Its components are spatially separated by 4.6 arcsec.
\item Two entries appear to be resolved WD--main-sequence pairs, one of them a WD--dM.  Both  pairs, \#'s 20 and 31, USNO-B 1053-0170642 and USNO-B 1251-0258748, are plotted in Fig.~\ref{fig:Fig5}d. Pair \# 20 is also plotted in Fig.~\ref{fig:Fig5}c, along with the WD component of pair \# 31.  A third pair, entry \#15, has a star flagged BLENDED, no deblend and is not plotted. We have retained the photometry in Table~\ref{tab:SDSS}, which, would place it in the WD region. Regardless of blending, it is difficult to envisage the object being in this region without it having a blue component.  Its luminous output is virtually flat over the \textit{u, g} and \textit{r} bands, indicative of a high temperature. Despite a faint output of about 21.8 mag, USNO-B 2nd epoch blue exposures show an extended object, aligned as per the position angles of Table~\ref{tab:list}. The average position angle alignment is within 14$\degr$ of that of the SDSS pair.
\item Two entries are possibly unresolved WD--dM binaries, as per \citet{Smolcic2004}, paired with main-sequence stars, although they are marginally in the WD--dM region and may in fact be main-sequence pairs. The apparent WD--dM components of both of these objects, \#'s 28a and 30a are plotted in Figs.~\ref{fig:Fig5}c \& d. 
\item Five entries are galaxies paired with main-sequence stars, one of the galaxies flagged, BLENDED, no deblend.
\item One entry may be an M giant/carbon paired with a main-sequence star as per \citet{Covey2007}. However, this designation is speculative and is based on its distance from the main-sequence locus.  Several of the dMs from the sample set, plotted in \ref{fig:Fig5}b are similarly outside of the locus, although not as wide. Both components are plotted in Fig.~\ref{fig:Fig5}d. This entry, \#33, USNO-B 1253-0259828, and one of the apparent WD--dM pairs, \#31, USNO-B 1251-0258748, exhibit the widest spectral separations. 
\end{enumerate} 

 \subsection{Discussion of SDSS results}

We emphasize that the identity of objects in Table~\ref{tab:SDSS} is tentative.  The sample set provided an illustration of how a cool WD, falling well within the colour--colour main-sequence locus could easily be misidentified.

 Our designation of two pairs as WD--dM compact binaries paired with main-sequence stars is open to question, as is the tentative identification of one pair as an M giant/carbon star paired with a main-sequence star.  It may well be that we have selected fourteen main-sequence binaries rather than eleven.  On the other hand the identification of two objects as WD--main-sequence pairings, and a possible third, appears to be credible. Though main-sequence pairs with extended USNO-B images were expected to be uncovered, the detection of three pairs closely spaced on the colour--colour plots is not easily explained. In all three instances one component is brighter than the other in all bands.

The alignment of all SDSS paired objects with the corresponding USNO-B displacement vector position angles stands out as the prime result of this exercise, and it goes a long way towards affirming our central premise regarding the detection of objects with displaced blue and red components. Though we have been clear throughout this paper in our preference for USNO-B entries with five observations as contrasted with those with four, we point out that eight of those aligned paired objects were entries with four observations.  
 
 \section{Conclusions}
\label{sec:Conclusions}

This feasibility study aimed at demonstrating the value of the USNO-B1.0 catalogue for selecting objects with displaced blue and red components. The specifics of the approach were developed with the aid of analysis of three sample test sets of known properties.  Over 91,000 USNO-B entries were then sifted and analysed and the results visually screened to obtain a candidate list of 146 objects. Slightly less than a quarter of this list coincided with SDSS data, which was used for tentative identification of object types and, most importantly, confirmed that displacement vectors formed from the USNO-B data aligned with corresponding SDSS pair position angles.

That confirmation leads us to reexamine restrictions that were added after the test set analysis. In particular, USNO-B five observation entries were required to have all displacement vector magnitudes greater than 0.35 arcsec and four observation entries were required to have vector magnitudes greater than 0.45 arcsec.  Relaxation of these strictures to lower thresholds would lead to a greater number of candidates and improved sensitivity at shorter pair separation distances.  Too great a relaxation will result in large numbers of false positives.  We expect that values of 0.3 arcsec for five observation entries, and 0.4 arcsec for four observation entries would be suitable. 

Another restriction that bears reevaluation is the maximum acceptable magnitude for \textit{B2}, which was set at 19.5~mag.  There was good reason for this restriction, especially for four observation entries, due to the dependence of two vectors on \textit{B2} and hence an interdependence between the two that is especially damaging for the errant values of \textit{B2} more likely to occur at faint magnitudes. However, this constraint was one of the most limiting with regard to reducing the number of entries suitable for analysis.  For those who would make use of this procedure, and specifically at the higher galactic latitudes, raising the ceiling to 20~mag is worthy of exploration. Our overwhelmingly negative result with three observation entries leads us to conclude that it is unproductive to pursue this course any further, though conceding that this conclusion is based on a small sample.  What would be worthwhile is the introduction of mixed epoch vectors (see Section~\ref{sec:Mixed epoch vectors}), which would expand the number of entries open to analysis. 

Although we have established the effectiveness of our approach in selecting objects with displaced coloration there is still the question of its usefulness. Those studying low mass ratio main-sequence binaries might find the efficiency marginal, and those seeking WD binaries would hardly be tempted by the low yield. And while we have clear evidence regarding the detection of WD and main-sequence pairs, our suppositions regarding the types of single object galaxies detected remains speculative until confirmed, or otherwise, by observation.  It seems clear that mounting an effective programme to exploit this tool would require a collaborative effort in which the different components of the team would have diverse interests. The motivation for such an effort assuredly lies in the expanse of USNO-B. This study has explored segments of the catalogue that, in total, encompass less than 0.01 per cent of that expanse.

\section*{Acknowledgements}

The author thanks David G. Monet for providing a copy of the USNO-B1.0 survey logs. He further thanks the referee, Dimitri Pourbaix, for his constructive comments and suggestions.

This research has made extensive use of the USNO Image and Catalogue Archive operated by the United States Naval Observatory, Flagstaff Station ( http://www.nofs.navy.mil/data/fchpix/ ).  Substantial use has also been made of the Simbad data base and the VizieR catalogue access tool, CDS, Strasbourg, France.  The original description of the Simbad and VizieR services were published, respectively, in A \& AS, 2000, 143, 9 and 23. 

Funding for SDSS-III has been provided by the Alfred P. Sloan Foundation, the Participating Institutions, the National Science Foundation, and the U.S. Department of Energy Office of Science.  The SDSS-III web site is http://www.sdss3.org/ .

SDSS-III is managed by the Astrophysical Research Consortium for the Participating Institutions of the SDSS-III Collaboration including the University of Arizona, the Brazilian Participation Group, Brookhaven National Laboratory, Carnegie Mellon University, University of Florida, the French Participation Group, the German Participation Group, Harvard University, the Instituto de Astrofisica de Canarias, the Michigan State/Notre Dame/JINA Participation Group, John Hopkins University, Lawrence Berkeley National Laboratory, Max Planck Institute for Astrophysics, Max Planck Institute for Extraterrestrial Physics, New Mexico University, New York University, Ohio State University, Pennsylvania State University, University of Portsmouth, Princeton University, the Spanish Participation Group, University of Tokyo, University of Utah, Vanderbilt University, University of Virginia, University of Washington, and Yale University.

\appendix

\section {Tables A1, A2, A3}
\label{sec:tables}
\suppressfloats[t]
\raggedbottom
 \begin{table}
 \begin{threeparttable}
	\centering
	\caption{Single star sample set. Coordinates are in equinox J2000, epoch J2000. }
	\label{tab:singlestar}
	\begin{tabular}{llccll}
		\hline
		& & RA & Dec & &\\
		USNO-B1.0 id. & Alt. id. & ($^{h}$ $^{m}$ $\overset{^s}{.}$) & ($\degr$ $'$ $\overset{''}{.} $)& \textit{r$_{rms}$} &pts.\\
		\hline
		 0824-0001341 & GJ 1002 & 00 06 43.13 & -07 32 16.87 & 0.14 & 2 \\ 
	1192-0001921 & GJ 1003 & 00 07 26.70 & +29 14 32.80 & 0.09 & 2 \\
	0833-0006357 & GJ 1012 & 00 28 39.46 & -06 39 49.08 & 0.11 & 0 \\ 
	0841-0005187 & GJ 1013 & 00 31 35.40 & -05 52 12.99 & 0.19 & 2 \\ 
	0855-0009698 & GJ 1025 & 01 00 56.37 & -04 26 56.33 & 0.15 & 1 \\ 
	1523-0040804 & GJ 51   & 01 03 19.87 & +62 21 55.93 & 0.13 & 4 \\ 
	0611-0011923 & GJ 1031 & 01 08 18.28 & -28 48 20.09 & 0.13 & 2 \\ 
	0654-0012495 & GJ 2021 & 01 09 18.71 & -24 30 23.49 & 0.15 & 3 \\ 
	0949-0011332 & GJ 3078 & 01 11 57.99 & +04 54 12.57 & 0.07 & 0 \\ 
	0838-0017238 & GJ 3119 & 01 51 04.10 & -06 07 05.15 & 0.12 & 4 \\ 
	0811-0020420 & LP709-43 & 02 10 03.65 & -08 52 59.62 & 0.20 & 3 \\ 
	0860-0020514 & LHS 1363	& 02 14 12.58 & -03 57 43.54 & 0.20 & 0 \\ 
	1068-0028941 & GAT 1370	& 02 53 00.88 & +16 52 52.76 & 0.12 & 2 \\ 
	0696-0041248 & LHS 1491	& 03 04 04.48 & -20 22 43.23 & 0.19 & 3 \\ 
	0592-0035569 & LP888-18	& 03 31 30.24 & -30 42 38.81 & 0.28 & 5 \\ 
	0839-0036868 & GJ 1065 	& 03 50 44.31 & -06 05 41.61 & 0.12 & 0 \\ 
	1070-0040309 & GJ 3253 	& 03 52 41.75 & +17 01 04.19 & 0.12 & 4 \\ 
	0655-0044012 & LHS 1630	& 04 07 20.48 & -24 29 13.53 & 0.07 & 1 \\ 
	0953-0073703 & GJ 1087 	& 05 56 25.47 & +05 21 48.57 & 0.13 & 2 \\ 
	1134-0116117 & GJ 232  & 06 24 41.28 & +23 25 59.08 & 0.13 & 5 \\ 
	1523-0188659 & GJ 3417 & 06 57 57.13 & +62 19 19.54 & 0.16 & 3 \\ 
	1427-0213945 & GJ 3421 & 07 03 55.76 & +52 42 06.89 & 0.22 & 2 \\ 
	1494-0183213 & GJ 3512 & 08 41 20.12 & +59 29 50.72 & 0.17 & 1 \\ 
	0865-0193882 & GJ 3517 & 08 53 36.18 & -03 29 32.29 & 0.22 & 3 \\ 
	1383-0239712 & GJ 1151 & 11 50 57.74 & +48 22 38.77 & 0.18 & 5 \\
	0662-0267348 & Note 1  & 12 14 08.71 & -23 45 16.79 & 0.14 & 2 \\ 
	0906-0208882 & GJ 1154 & 12 14 16.56 & +00 37 26.47 & 0.16 & 3 \\ 
	0956-0217859 & GJ 493.1 & 13 00 33.54 & +05 41 08.21 & 0.13 & 0 \\ 
	1497-0221149 & GJ 3855 & 14 30 37.79 & +59 43 25.15 & 0.13 & 2 \\ 
	1308-0269841 & GJ 3959 & 16 31 18.79 & +40 51 51.93 & 0.07 & 0 \\ 
	1124-0308674 & GJ 3976 & 16 50 57.95 & +22 27 05.68 & 0.09 & 1 \\ 
	0856-0301315 & GJ 1207 & 16 57 05.72 & -04 20 56.64 & 0.10 & -1 \\ 
	1414-0271453 & GJ 3988 & 17 03 23.90 & +51 24 23.03 & 0.12 & 3 \\ 
	1608-0117171 & GJ 1221 & 17 48 00.00 & +70 52 36.17 & 0.16  & 4 \\ 
	1542-0210205 & G 227-22	& 18 02 16.63 & +64 15 44.62 & 0.08 & 1 \\ 
	1520-0275071 & GJ 1227 & 18 22 27.14 & +62 03 01.95 & 0.22 & 3 \\ 
	1054-0588142 & GJ 1256 & 20 40 33.86 & +15 29 58.89 & 0.13 & 3 \\ 
	0723-1162500 & GJ 4274 & 22 23 06.99 & -17 36 26.29 & 0.13 & 1 \\ 
	1341-0523884 & GJ 905  & 23 41 55.04 & +44 10 39.07 & 0.17 & 1 \\ 
	0738-0841136 & GJ 4360 & 23 45 31.25 & -16 10 20.07 & 0.10 & 1 \\ 

		\hline
	\end{tabular}
	\begin{tablenotes}
	\small
	\item Note 1. Alternate identification is 2MASS J12140866-2345172
	\end{tablenotes}
	\end{threeparttable}
\end{table}

\raggedbottom
 \begin{table*}
	\centering
	\caption{Binary sample sets. Coordinates are in equinox J2000, epoch J2000. For both sets of data only one component of the binary is listed under `Alternate ident.'  The references for the dM binary set are \citet{Janson2012,Janson2014}.  References for the WD--dM binary set (in parentheses after separation) are: (1) \citet{Hoard2007}, (2) \citet*{Silvestri2005} and (3) \citet{Farihi2006,Farihi2010}. `WDS' indicates that the separation was obtained from the USNO Washington Double Star Catalogue. The SDSS column indicates whether or not (Y or blank) SDSS data is available for the WD--dM  binary entry (see Section~\ref{sec:Binary test set}.}
	\label{tab:binaries}
	\begin{tabular}{ccccccc}
	\hline
	& & & & Separation & SDSS &\\
		USNO-B1.0 ident. & Alternate ident. & RA ($^{h}$ $^{m}$ $\overset{^s}{.}$)  & Dec ($\degr$ $ ' $ $\overset{''}{.} $) & (arcsec) & (WD--dM) & Point count\\
	\hline
	& & \multicolumn{2}{c}{dM binary sample set}\\
	& 2MASS 	& & & & &\\
	& & & & & &\\
	0854-0005123 & J00325313-0434068 & 00 32 53.13 & -04 34 06.9 & 0.42-0.51  &  & 1\\ 
1308-0018556 & J01034210+4041158 & 01 03 42.11 & +40 51 16.1 & 1.5-2.5   &  & 2\\
0823-0016981 & J01132958-0738088 & 01 13 29.59 & -07 38 08.8 & 2.9    &  & 3\\
0832-0024712 & J01365516-0647379 & 01 36 55.18 & -06 47 38.5 & 5.4    &  & 1\\
0750-0018344 & J01535076-1459503 & 01 53 50.76 & -14 59 50.2 & 2.8    &  & 3\\
0805-0022317 & J02155892-0929121 & 02 15 58.93 & -09 29 12.0 & 0.06/3.4  &  & 4\\
0666-0026418 & J02165488-2322133 & 02 16 54.80 & -23 22 12.0 & 4.3    &  & 0\\
0605-0025977 & J02271603-2929263 & 02 27 16.04 & -29 29 26.2 & 1.9    &  & 3\\
0883-0068712 & J04132663-0139211 & 04 13 26.62 & -01 39 21.2 & 0.74    &  & 3\\
0663-0056479 & J05100488-2340148 & 05 10 04.88 & -23 40 14.9 & 1.8    &  & 2\\
0890-0056913 & J05191382-0059423 & 05 19 13.80 & -00 59 43.7 & 1.1    &  & 5\\
0573-0081113 & J05344858-3239362 & 05 34 48.59 & -32 39 36.3 & 2.1/0.44  &  & 0\\
0848-0062045 & J05350429-0508125 & 05 35 04.48 & -05 08 13.3 & 4.1    &  & 2\\
0820-0079747 & J05464932-0757427 & 05 46 49.34 & -07 57 44.2 & 2.8    &  & 4\\
0780-0096591 & J05494272-1158500 & 05 49 42.60 & -11 58 49.7 & 3.4    &  & 0\\
0780-0096591 & J06281861-0110504 & 06 28 18.59 & -01 10 50.7 & 1.4    &  & 5\\
1377-0208428 & J06351837+4745366 & 06 35 18.64 & +47 45 37.2 & 3.5$\sim$3.9   &  & 2\\
0696-0115277 & J06583980-2021526 & 06 58 39.76 & -20 21 52.9 & 1.4    &  & 4\\
1344-0192252 & J07505369+4428181 & 07 50 53.69 & +44 28 18.5 & 2.0    &  & 1\\
1302-0190151 & J08310177+4012115 & 08 31 01.78 & +40 12 11.8 & 1.9    &  & 2\\
0761-0200741 & J08483645-1353083 & 08 48 36.72 & -13 53 07.7 & 7.4/0.3   &  & 5\\
0591-0203730 & J08540240-3051366 & 08 54 02.38 & -30 51 36.7 & 1.7    &  & 2\\
0885-0192790 & J10122171-0128160 & 10 12 21.68 & -01 28 15.9 & 2.9    &  & 2\\
1053-0192207 & J10364483+1521394 & 10 36 44.84 & +15 21 39.3 & 0.17/1.0  &  & 0\\
1216-0196987 & J11281625+3136017 & 11 28 16.24 & +31 36 01.8 & 1.1    &  & 1\\
0786-0233601 & J12134173-1122405 & 12 13 41.73 & -11 22 40.9 & 1.3    &  & 3\\
1326-0292758 & J13015919+4241160 & 13 01 59.17 & +42 41 16.3 & 2.9    &  & 1\\
1222-0276082 & J13120525+3213332 & 13 12 05.30 & +32 13 33.5 & 0.9    &  & 0\\
1548-0176179 & J14190331+6451463 & 14 19 03.32 & +64 51 46.7 & 4.6    &  & 2\\
1035-0224299 & J14360274+1334484 & 14 36 02.73 & +13 34 48.6 & 1.2     &  & 1\\
1341-0273828 & J14450627+4409393 & 14 45 06.29 & +44 09 39.2 & 5.2    &  & 2\\
0893-0244088 & J15032251-0040310 & 15 03 22.65 & -00 40 30.9 & 3.8    &  & 1\\
1252-0234146 & J15553178+3512028 & 15 55 31.80 & +35 12 02.7 & 1.5-1.6   &  & 4\\
1340-0271823 & J15594729+4403595 & 15 59 47.31 & +44 03 59.9 & 5.6    &  & 4\\
1538-0209444 & J16363309+6353452 & 16 36 33.05 & +63 53 44.9 & 0.2/3.4   &  & 2\\
1317-0302575 & J16460779+4142057 & 16 46 07.73 & +41 42 07.4 & 5.7    &  & 4\\
1259-0251199 & J16510995+3555071 & 16 51 09.97 & +35 55 07.2 & 1.0    &  & 2\\
1023-0718446 & J21035992+1218570 & 21 03 59.93 & +12 18 57.0 & 0.83    &  & -1\\
0817-0690783 & J21091375-0814041 & 21 09 13.74 & -08 14 03.9 & 0.97    &  & 1\\
0803-0681472 & J22332264-0936537 & 22 33 22.65 & -09 36 53.5 & 1.5    &  & 2\\

\hline	
\end{tabular}
\end{table*}

\raggedbottom
\begin{table*}
	\centering
	\contcaption{binaries}
	\begin{tabular}{ccccccc} 
	\hline
	& & & & Sep. (Ref.) & SDSS &\\
		USNO-B1.0 ident. & Alternate ident. & RA ($^{h}$ $^{m}$ $\overset{^s}{.}$)  & Dec ($\degr$ $ ' $ $ \overset{''}{.} $) & (arcsec) & (WD--dM) & Point count\\
\hline
& & & & & &\\		
	& & & & & &\\
	& & \multicolumn{2}{c}{WD--dM binary sample set}\\
	& WD ident. & & & Sep. (Ref.)& &\\
	& & & & & &\\
1000-0002069 & WD 0014+097 & 00 16 56.15 & +10 03 58.8 & <4 (1) & Y & 6\\ 
0742-0019772 & WD 0158-160 & 02 00 56.69 & -15 46 10.7 & 5.7-7.8/WDS (2) & Y & 1\\ 
1036-0021298 & WD 0205+133 & 02 08 03.49 & +13 36 25.1 & 1.257 (3) & Y & 7\\ 
1277-0045811 & WD 0217+375 & 02 20 25.25 & +37 47 30.9 & $\sim$2 (2) &  &  2\\ 
0896-0030878 & WD 0257-005 & 03 00 24.56 & -00 23 41.8 & 0.978 (3) & Y & 4\\ 
1549-0145856 & WD 0911+651 & 09 15 32.05 & +64 56 41.2 & 6 (2) & Y & 5\\ 
1098-0172007 & WD 0915+201 & 09 18 33.04 & +19 53 07.8 & 2.312 (3) & Y & 6\\ 
0923-0231599 & WD 0933+025 & 09 35 40.70 & +02 22 00.2 & 1.232 (3) & Y & 6\\ 
1349-0213613 & WD 0949+451 & 09 52 22.00 & +44 54 29.8 & 2.892 (3) & Y & 5\\ 
1083-0198558 & WD 0950+185 & 09 52 45.80 & +18 21 03.1 & 1.1 (1) & Y & 7\\ 
0943-0177699 & WD 0956+045 & 09 58 37.25 & +04 21 31.0 & 2 (1) & Y & 4\\ 
0941-0190204 & WD 1104+044 & 11 07 04.76 & +04 09 09.2 & $\sim$3 (2) & Y & 7\\ 
1213-0187197 & WD 1106+316 & 11 08 43.05 & +31 23 56.2 & 0.478 (3) & Y & 2\\ 
1086-0195362 & WD 1123+189 & 11 26 19.09 & +18 39 17.2 & 1.3 (1) & Y & 5\\ 
1026-0245938 & WD 1157+129 & 11 59 15.65 & +12 39 30.0 & 0.564 (3) & Y & 3\\ 
1361-0222911 & WD 1210+464 & 12 12 59.64 & +46 09 47.0 & 1.043 (3) & Y & 7\\ 
0929-0275178 & WD 1214+032 & 12 16 51.86 & +02 58 04.7 & 2 (1) & Y & 6\\ 
1220-0233210 & WD 1215+322 & 12 17 33.26 & +32 05 09.0 & 6.6-7/WDS (2) & Y & 4\\ 
0893-0216590 & WD 1236-004 & 12 38 36.34 & -00 40 42.2 & 0.658 (3) & Y & 4\\
1651-0066694 & WD 1240+754 & 12 42 03.85 & +75 08 43.8 & 6 (2) &  &  5\\ 
0755-0271141 & WD 1307-141 & 13 10 22.53 & -14 27 09.1 & 2.133 (3) &  & 7\\ 
1384-0238519 & WD 1333+487 & 13 36 01.95 & -48 28 46.7 & 2.947 (3) & Y & 6\\ 
1403-0256955 & WD 1402+506 & 14 04 08.95 & +50 20 38.4 & 5 (2) &  Y & 6\\ 
1474-0288101 & WD 1419+576 & 14 21 05.37 & +57 24 57.4 & 0.655 (3) & Y & 4\\ 
1268-0246415 & WD 1435+370 & 14 37 36.71 & +36 51 37.5 & 1.251 (3) & Y & 6\\ 
0591-0203730 & WD 1443+337 & 14 46 00.72 & +33 28 50.1 & 0.679 (3) & Y & 6\\ 
1247-0222694 & WD 1502+349 & 15 04 31.86 & +34 47 00.0 & 1.913 (3) & Y & 5\\ 
0760-0301587 & WD 1539-137 & 15 42 00.94 & -13 56 06.6 & $\sim$7    (2) &  & 5\\
1515-0219383 & WD 1558+616 & 15 58 55.55 & +61 32 03.8 & 0.72  (3) & Y & 6\\ 
1423-0310703 & WD 1619+525 & 16 20 24.39 & +52 23 21.5 & 2.596/0.466 (3) & Y & 4\\ 
1545-0210208 & WD 1833+644 & 18 33 29.22 & +64 31 52.2 & 0.079/1.82 (3) & Y & 2\\ 
1386-0457478 & WD 2224+483 & 22 26 16.63 & +48 37 31.1 & $\sim$2 (2) &  &  3\\ 
0765-0711730 & WD 2318-137 & 23 21 14.41 & -13 27 38.4 & $\sim$3 (1) &  &  3\\ 
0738-0840954 & WD 2341-164 & 23 44 20.20 & -16 10 50.1 & 6.2-6.7/WDS (2) &  & 7\\ 
1165-0040920 & WD 0325+263 & 03 28 03.68 & +26 31 52.4 & 6-6.3/WDS (2) &  & 6\\
0667-0042912 & WD 0357-233 & 03 59 04.89 & -23 12 25.3 & 1.19  (3) &  & 6\\ 
0580-0061995 & WD 0358-321 & 04 00 03.19 & -31 57 53.0 & $\sim$10 (2) &  & 1\\ 
1726-0023464 & WD 0725+827 & 07 36 32.91 & +82 36 46.9 & $\sim$5 (2) & Y & 0\\ 
1187-0163669 & WD 0824+288 & 08 27 05.14 & +28 44 02.7 & 0.08/3.33 (3) & Y & 7\\ 
1394-0199142 & WD 0842+496 & 08 46 23.43 & +49 25 36.7 & 5 (2) & Y & 4\\ 
	
	\hline
	\end{tabular}
\end{table*}

\raggedbottom
\begin{table*}
	\centering
	\caption{Candidate list. Entries are arranged by increasing order of the USNO-B1.0 prefix and, within each prefix, by increasing RA.  The prefix indicates the distance from the southern pole in increments of 0.1$\degr$.  The last column provides a subjective assessment of the USNO-B1.0 images, either compact (c) or extended (e).  Ambiguous instances have been left blank.  Multiple objects on the images are noted where there is a possibility of confused readings. The last column further indicates availability of SDSS data.}
	\label{tab:list}
	\begin{tabular}{ccccccccccc} 
	\hline
	& eqnx. J2000, epoch J2000 & No. & PA (${\degr}$) & PA (${\degr}$)  & PA (${\degr}$)   & Vec. mag. & Vec. mag. & Vec. mag. &	USNO-B1.0 \\
	USNO-B1.0 & ($^{h}$ $^{m}$ $\overset{^s}{.}$ $\pm$ $\degr$ $ ' $ $ \overset{''}{.} $) & obs. & \textbf{R1-B1} & \textbf{R2-B2} & \textbf{I2-B2} & \textbf{R1-B1} &  \textbf{R2-B2} & \textbf{I2-B2} & images,\\
	& & & & & & (arcsec) & (arcsec) & (arcsec) & SDSS data\\
	\hline
	0034-0006865 & 03 34 23.39 -86 31 53.87   & 4 &  & 284.0 & 266.4 &  & 0.45 & 0.64 & c	\\
0034-0006916 & 03 36 18.42 -86 32 05.82   & 4 &  & 11.3 & 36.4 &  & 0.46 & 0.47 & c	\\
0035-0006480 & 03 26 04.95 -86 26 02.40   & 4 &  & 312.0 & 305.4 &  & 0.69 & 0.76 & c	\\
0035-0006483 & 03 26 24.16 -86 24 35.65   & 4 &  & 223.3 & 209.2 &  & 0.47 & 0.57 & c	\\
0035-0006625 & 03 30 36.61 -86 29 28.38   & 4 &  & 185.5 & 189.6 &  & 0.83 & 1.02 & c	\\
0035-0006702 & 03 32 54.71 -86 29 44.03   & 4 &  & 206.1 & 226.1 &  & 0.52 & 0.72 & c	\\
0035-0007013 & 03 44 10.02 -86 24 51.29   & 4 &  & 136.6 & 110.2 &  & 0.51 & 0.84 & c	\\
0036-0007224 & 03 23 50.75 -86 20 56.59   & 4 &  & 71.9 & 65.4 &  & 0.48 & 1.06 & \\
0036-0007247 & 03 24 33.90 -86 22 34.49   & 4 &  & 240.9 & 228.0 &  & 0.51 & 1.36 & \\
0036-0007724 & 03 39 21.95 -86 20 38.25   & 4 &  & 231.5 & 238.8 &  & 0.50 & 0.50 & c	\\
0036-0008100 & 03 51 14.50 -86 18 37.49   & 4 &  & 89.0 & 72.3 &  & 0.56 & 1.05 & e	\\
0037-0007134 & 03 20 47.13 -86 15 51.75   & 4 &  & 176.3 & 167.7 &  & 0.46 & 0.56 & c	\\
0037-0007263 & 03 25 07.19 -86 14 03.36   & 4 &  & 195.5 & 193.4 &  & 0.49 & 0.69 & c	\\
0037-0007345 & 03 27 42.47 -86 14 58.75   & 4 &  & 267.6 & 277.9 &  & 0.47 & 0.80 & c	\\
0038-0007574 & 03 37 31.44 -86 07 36.70   & 4 &  & 323.0 & 327.7 &  & 0.66 & 0.67 & c	\\
0039-0006764 & 03 22 07.96 -86 03 28.69   & 4 &  & 248.4 & 278.0 &  & 0.46 & 0.58 & e	\\
0039-0006822 & 03 24 21.95 -86 02 16.30   & 4 &  & 43.0 & 45.5 &  & 1.03 & 1.68 & \\
0039-0007090 & 03 33 13.88 -86 02 13.49   & 4 &  & 236.9 & 231.7 &  & 0.59 & 0.85 & e	\\
0039-0007209 & 03 37 18.24 -86 01 33.75   & 4 &  & 225.0 & 220.4 &  & 0.45 & 0.80 & c	\\
0039-0007813 & 03 55 27.13 -86 02 53.23   & 4 &  & 216.5 & 230.2 &  & 0.67 & 0.78 & c	\\
0040-0007055 & 03 23 08.83 -85 56 06.63   & 4 &  & 31.7 & 21.4 &  & 0.80 & 0.99 & \\
0040-0007079 & 03 23 48.42 -85 55 50.45   & 4 &  & 43.9 & 50.9 &  & 1.12 & 1.24 & \\
0040-0007440 & 03 34 40.62 -85 54 25.42   & 4 &  & 222.3 & 199.6 &  & 0.59 & 0.96 & e	\\
0040-0007463 & 03 35 41.76 -85 59 36.79   & 4 &  & 46.3 & 38.0 &  & 0.61 & 1.22 & e	\\
0040-0007494 & 03 36 23.31 -85 56 55.05   & 4 &  & 35.2 & 42.0 &  & 0.54 & 0.82 & e	\\
0040-0007607 & 03 39 48.88 -85 55 19.06   & 4 &  & 321.7 & 332.3 &  & 0.55 & 0.69 & e	\\
0040-0007650 & 03 41 12.66 -85 59 47.48   & 4 &  & 174.9 & 191.9 &  & 0.56 & 0.58 & e	\\
0040-0008096 & 03 53 03.94 -85 57 58.69   & 4 &  & 81.0 & 65.2 &  & 0.58 & 1.00 & e	\\
0041-0007489 & 03 27 51.71 -85 49 50.45   & 4 &  & 189.3 & 173.1 &  & 0.56 & 0.58 & c	\\
0041-0007524 & 03 29 05.75 -85 52 51.31   & 4 &  & 207.1 & 194.5 &  & 0.48 & 0.84 & c	\\
0041-0007540 & 03 29 56.67 -85 50 38.37   & 4 &  & 203.3 & 197.1 &  & 0.78 & 0.82 & e	\\
0041-0007589 & 03 31 17.98 -85 53 32.22   & 4 &  & 70.8 & 95.9 &  & 0.46 & 0.58 & \\
0041-0007670 & 03 34 18.60 -85 52 26.38   & 4 &  & 248.2 & 249.0 &  & 0.54 & 0.56 & e	\\
0041-0007801 & 03 38 16.01 -85 48 58.35   & 4 &  & 35.0 & 51.5 &  & 0.49 & 0.63 & c	\\
0041-0007853 & 03 40 04.92 -85 50 34.61   & 4 &  & 185.0 & 207.1 &  & 0.57 & 0.92 & c	\\
0041-0007916 & 03 41 30.37 -85 49 22.88   & 4 &  & 241.2 & 273.1 &  & 0.46 & 0.56 & e.	\\
0041-0007992 & 03 43 34.17 -85 49 36.71   & 4 &  & 297.7 & 271.1 &  & 0.45 & 0.50 & c	\\
0042-0007934 & 03 30 52.83 -85 47 48.94   & 4 &  & 122.7 & 121.6 &  & 0.50 & 0.67 & c	\\
0042-0008005 & 03 33 40.39 -85 46 42.21   & 4 &  & 353.7 & 0.0 &  & 0.82 & 0.92 & e	\\
0042-0008079 & 03 35 51.82 -85 47 16.22   & 4 &  & 69.7 & 60.0 &  & 0.90 & 1.04 & c	\\
0042-0008136 & 03 37 30.59 -85 46 18.71   & 4 &  & 64.4 & 84.9 &  & 0.51 & 0.56 & e	\\
0042-0008199 & 03 39 44.20 -85 45 59.74   & 4 &  & 163.4 & 186.8 &  & 0.59 & 0.67 & e	\\
0500-0838052 & 22 38 28.84 -39 54 45.28   & 4 &  & 139.6 & 148.2 &  & 0.62 & 1.80 & e	\\
0501-0828963 & 22 39 24.19 -39 50 11.80   & 4 &  & 199.9 & 198.2 &  & 0.62 & 0.67 & c	\\
0502-0839757 & 22 39 24.45 -39 43 00.74   & 4 &  & 288.2 & 310.0 &  & 0.64 & 0.89 & e	\\
0503-0821145 & 22 39 30.94 -39 39 54.60   & 4 &  & 54.9 & 34.3 &  & 0.66 & 1.03 & c	\\
0503-0821243 & 22 40 8.81 - 39 40 50.99   & 4 &  & 277.7 & 290.6 &  & 0.52 & 0.82 & c	\\
0504-0831027 & 22 38 21.38 -39 35 35.90   & 4 &  & 306.0 & 319.8 &  & 0.68 & 1.43 & e	\\
0504-0831304 & 22 40 08.99 -39 31 57.99   & 4 &  & 233.9 & 240.4 &  & 0.46 & 0.59 & c	\\
0505-0818836 & 22 38 15.94 -39 26 49.66   & 4 &  & 309.6 & 330.5 &  & 0.45 & 0.53 & c	\\
0505-0818918 & 22 38 37.35 -39 29 18.48   & 4 &  & 357.9 & 13.0 &  & 0.54 & 0.80 & c	\\
0505-0819180 & 22 40 15.88 -39 27 42.32   & 4 &  & 294.9 & 264.6 &  & 0.45 & 0.53 & e	\\
0507-0816170 & 22 39 18.00 -39 14 04.60   & 4 &  & 12.3 & 353.5 &  & 0.47 & 0.53 & c	\\
0507-0816211 & 22 39 33.85 -39 15 00.52   & 4 &  & 6.8 & 4.3 &  & 0.59 & 0.66 & c	\\
0507-0816304 & 22 39 59.28 -39 16 17.55   & 4 &  & 70.4 & 67.1 &  & 0.48 & 0.56 & c	\\
0507-0816437 & 22 40 48.17 -39 13 37.22   & 4 &  & 276.2 & 294.0 &  & 0.46 & 0.49 & c	\\
0527-0283790 & 10 06 49.87 -37 17 24.46   & 4 &  & 142.8 & 136.8 &  & 0.68 & 0.69 & e	\\
0528-0283777 & 10 06 02.02 -37 07 40.82   & 4 &  & 93.4 & 100.7 &  & 0.67 & 0.75 & e	\\
\hline	
\end{tabular}
\end{table*}

\raggedbottom
\begin{table*}
	\centering
	\contcaption{Candidate list.}
	\begin{tabular}{ccccccccccc} 
	\hline
	& eqnx. J2000, epoch J2000 & No. & PA (${\degr}$) & PA (${\degr}$) & PA (${\degr}$)  & Vec. mag. & Vec. mag. & Vec. mag. &	USNO-B1.0 \\
	USNO-B1.0 &  ($^{h}$ $^{m}$ $\overset{^s}{.}$ $\pm$ $\degr$ $ ' $ $ \overset{''}{.} $)  & obs. & \textbf{R1-B1} & \textbf{R2-B2} & \textbf{I2-B2} & \textbf{R1-B1} &  \textbf{R2-B2} & \textbf{I2-B2} & images,\\
	& & & & & & (arcsec) & (arcsec) & (arcsec) & SDSS data\\
	\hline
0528-0283958 & 10 06 28.75 -37 11 57.48   & 4 &  & 316.6 & 309.5 &  & 2.04 & 2.86 & e	\\
0528-0284188 & 10 06 58.73 -37 09 47.04   & 4 &  & 187.3 & 166.4 &  & 0.47 & 0.60 & c	\\
0528-0284338 & 10 07 14.49 -37 11 15.04   & 4 &  & 343.1 & 320.3 &  & 0.48 & 0.61 & e	\\
0529-0277001 & 10 05 54.80 -37 02 48.15   & 4 &  & 217.1 & 225.8 &  & 0.83 & 1.03 & e	\\
0529-0277033 & 10 05 58.06 -37 05 53.39   & 4 &  & 270.5 & 260.0 &  & 1.17 & 1.78 & e	\\
0529-0277044 & 10 05 59.77 -37 03 16.30   & 4 &  & 107.3 & 97.5 &  & 0.47 & 0.53 & e	\\
0529-0277813 & 10 07 42.22 -37 04 55.50   & 4 &  & 214.6 & 219.4 &  & 0.51 & 0.65 & e	\\
0530-0039917 & 03 07 46.17 -36 59 29.24   & 4 &  & 29.1 & 26.9 &  & 0.70 & 0.73 & c	\\
0531-0039963 & 03 06 59.88 -36 51 06.92   & 4 &  & 58.4 & 67.9 &  & 0.76 & 1.78 & e	\\
0534-0033956 & 03 04 57.23 -36 33 15.53   & 4 &  & 135.9 & 167.1 &  & 0.47 & 0.49 & c	\\
0535-0026245 & 03 05 19.05 -36 27 43.74   & 4 &  & 257.7 & 261.5 &  & 0.47 & 0.48 & c	\\
0536-0025111 & 03 06 36.93 -36 22 00.77   & 4 &  & 358.8 & 355.1 &  & 0.47 & 0.47 & e	\\
0571-0375276 & 12 40 00.91 -32 48 03.75   & 4 &  & 51.0 & 48.5 &  & 0.48 & 0.81 & e	\\
0572-0382482 & 12 40 26.13 -32 43 59.17   & 4 &  & 112.1 & 112.9 &  & 0.69 & 1.77 & e	\\
0574-0375910 & 12 39 51.13 -32 31 56.00   & 4 &  & 314.3 & 312.7 &  & 1.16 & 1.22 & e	\\
0574-0376072 & 12 40 20.12 -32 35 23.90   & 5 & 344.5 & 330.9 & 347.6 & 0.98 & 0.41 & 0.42 & c\\
0630-0013411 & 00 52 16.45 -26 56 02.09   & 4 &  & 117.5 & 112.8 &  & 0.54 & 0.54 & c	\\
0814-0341014 & 17 24 40.51 -08 34 28.42   & 5 & 293.4 & 314.1 & 300.6 & 0.48 & 0.45 & 1.53 & e\\
0814-0341131 & 17 24 49.97 -08 30 06.85   & 4 & 273.0 &  & 272.1 & 0.58 &  & 0.82 & c	\\
0815-0344448 & 17 24 16.84 -08 24 39.76   & 5 & 284.6 & 287.8 & 288.9 & 1.23 & 1.61 & 2.01 & c\\
0816-0364222 & 17 24 06.14 -08 21 31.70   & 5 & 250.2 & 254.1 & 276.6 & 0.80 & 0.44 & 1.30 & c\\
0816-0364339 & 17 24 15.05 -08 19 11.23   & 5 & 170.1 & 178.6 & 166.6 & 0.41 & 0.41 & 0.43 & c\\
0816-0364528 & 17 24 27.29 -08 21 42.19   & 5 & 168.0 & 175.9 & 185.2 & 0.82 & 0.83 & 1.09 & c\\
0816-0364695 & 17 24 39.93 -08 23 29.39   & 5 & 266.8 & 263.5 & 274.8 & 0.53 & 0.44 & 0.59 & c\\
0816-0364783 & 17 24 47.64 -08 21 34.04   & 5 & 59.2 & 55.0 & 34.6 & 0.49 & 0.49 & 0.55 & c	\\
0816-0364857 & 17 24 53.97 -08 21 38.94   & 5 & 256.0 & 264.6 & 271.7 & 0.58 & 1.17 & 1.04 & c\\
0816-0364905 & 17 24 58.50 -08 19 08.07   & 5 & 284.0 & 291.6 & 280.9 & 0.66 & 0.52 & 0.58 & c\\
0816-0364932 & 17 25 00.99 -08 21 36.13   & 4 & 281.2 & 266.7 &  & 0.72 & 0.69 &  & c	\\
0817-0389673 & 17 24 43.23 -08 16 01.28   & 5 & 196.1 & 182.4 & 213.3 & 0.40 & 0.47 & 0.73 & c\\
0924-0009909 & 00 44 22.53 +02 29 49.38   & 5 & 339.4 & 351.1 & 332.0 & 0.74 & 0.84 & 1.11 & e/ SDSS\\	
0966-0584913 & 21 27 51.05 +06 39 50.63   & 4 & 119.9 & 108.4 &  & 1.08 & 0.89 &  & e/SDSS\\
0969-0633655 & 21 27 36.75 +06 59 24.69   & 4 &  & 205.5 & 206.3 &  & 0.72 & 1.06 & c/SDSS\\
0969-0633980 & 21 28 26.31 +06 58 37.56   & 5 & 92.1 & 82.7 & 86.3 & 0.54 & 0.47 & 0.46 & e/  SDSS\\
0969-0634114 & 21 28 51.73 +06 57 47.38  & 5 & 204.7 & 195.0 & 174.9 & 0.67 & 0.73 & 0.90 & e/  SDSS\\
0970-0663419 & 21 28 14.66 +07 00 50.71   & 5 & 88.8 & 46.1 & 76.8 & 0.98 & 0.36 & 0.48 & c/SDSS\\
0971-0681235 & 21 27 26.09 +07 10 06.27   & 5 & 207.6 & 232.7 & 232.3 & 0.47 & 0.48 & 0.56 & c/e / SDSS\\
0971-0681479 & 21 28 02.38 +07 10 16.19   & 4 &  & 269.0 & 264.2 &  & 0.58 & 0.59 & c/e/SDSS\\
0971-0681705 & 21 28 36.68 +07 08 47.92   & 4 &  & 21.4 & 18.6 &  & 2.28 & 3.07 & mult. objs./  SDSS\\
0972-0698506 & 21 28 29.27 +07 16 14.50   & 5 & 188.2 & 214.6 & 242.9 & 1.33 & 0.35 & 0.44 & c/e/SDSS\\
0984-0266003 & 14 31 54.05 +08 25 59.85   & 4 &  & 192.5 & 211.1 &  & 0.46 & 0.68 & e/SDSS\\
1006-0190377 & 11 03 11.19 +10 39 32.61   & 5 & 0.9 & 358.8 & 5.7 & 0.62 & 0.48 & 0.40 & SDSS\\
1006-0190380 & 11 03 13.15 +10 37 53.11   & 4 &  & 340.4 & 311.0 &  & 0.48 & 0.61 & c/SDSS\\
1009-0036682 & 04 03 26.36 +10 57 33.52   & 4 & 16.4 & 43.5 &  & 0.64 & 0.55 &  & e\\
1009-0190583 & 11 03 05.92 +10 56 57.01   & 4 &  & 150.8 & 167.4 &  & 0.49 & 0.59 & c/SDSS\\
1009-0190822 & 11 04 35.74 +10 57 46.08   & 5 & 230.1 & 257.7 & 276.2 & 0.56 & 0.47 & 0.55 & e/SDSS\\
1010-0189647 & 11 03 17.04 +11 02 19.09   & 4 &  & 236.5 & 232.6 &  & 0.82 & 0.91 & c/SDSS\\
1010-0189899 & 11 04 57.61 +11 01 50.25   & 5 & 310.6 & 276.6 & 277.0 & 0.55 & 0.43 & 0.41 & c/SDSS\\
1011-0036127 & 04 02 26.56 +11 09 06.87   & 5 & 171.1 & 180.0 & 209.0 & 0.52 & 0.43 & 0.64 & c\\
1011-0036201 & 04 03 06.37 +11 08 42.15   & 4 & 249.2 & 263.2 &  & 1.29 & 1.10 &  & e	\\
1012-0035157 & 04 01 57.24 +11 15 07.65   & 4 &  & 303.7 & 297.9 &  & 0.47 & 0.58 & c	\\
1012-0035331 & 04 03 08.19 +11 15 47.12   & 4 &  & 256.0 & 268.9 &  & 0.45 & 0.52 & c	\\
1053-0170335 & 08 35 36.61 +15 19 47.67  & 5 & 142.0 & 151.9 & 139.1 & 1.04 & 0.49 & 0.60 & e/SDSS\\
1053-0170642 & 08 36 47.71 +15 22 10.66  & 5 & 280.4 & 246 .0 & 255.8 & 0.50 & 0.59 & 0.65 & e/SDSS\\
1054-0169679 & 08 35 39.14 +15 25 35.39 & 4 & & 312.2 & 309.9 & & 0.88 & 1.68 & e/SDSS\\
1055-0172337 & 08 37 22.32 +15 30 36.62 & 5 & 138.5 & 153.4 & 160.0 & 0.94 & 0.96 & 1.07 & e/SDSS\\
1058-0172600 & 08 36 25.34 +15 51 11.22 & 5 & 181.5 & 186.8 & 190.9 & 0.38 & 0.50 & 0.53 & e/SDSS\\
1058-0172658 & 08 36 50.58 +15 48 58.61 & 5 & 232.1 & 251.2 & 250.3 & 0.57 & 0.56 & 0.45 & e/SDSS\\
1108-0181990 & 09 33 44.39 +20 53 00.40 & 4 & & 268.1 & 274.4 & & 0.92 & 1.04 & e/SDSS\\
1169-0233621 & 12 51 33.52 +26 55 08.74   & 5 & 237.7 & 238.8 & 263.7 & 0.45 & 0.50 & 1.10 & e/SDSS\\
1170-0240143 & 12 50 34.32 +27 00 59.22   & 5 & 265.6 & 259.3 & 284.3 & 0.39 & 0.38 & 0.53 & c/SDSS\\
1188-0095610 & 05 45 12.60 +28 49 57.91   & 4 & 255.5 &  & 270.7 & 0.64 &  & 0.82 & e	\\
\hline	
	\end{tabular}
\end{table*}

\raggedbottom
\begin{table*}
	\centering
	\contcaption{Candidate list.}
	\begin{tabular}{ccccccccccc} 	
	\hline
	& eqnx. J2000, epoch J2000 & No. & PA (${\degr}$) & PA (${\degr}$)  & PA (${\degr}$)   & Vec. mag. & Vec. mag. & Vec. mag. &	USNO-B1.0 \\
	USNO-B1.0 &  ($^{h}$ $^{m}$ $\overset{^s}{.}$ $\pm$ $\degr$ $ ' $ $ \overset{''}{.} $)  & obs. & \textbf{R1-B1} & \textbf{R2-B2} & \textbf{I2-B2} & \textbf{R1-B1} &  \textbf{R2-B2} & \textbf{I2-B2} & images,\\
	& & & & & & (arcsec) & (arcsec) & (arcsec) & SDSS data\\
	\hline
1188-0095741 & 05 45 26.13 +28 53 05.66   & 5 & 183.7 & 150.5 & 165.6 & 0.46 & 0.53 & 0.76 & e\\
1188-0096051 & 05 46 01.32 +28 50 50.21   & 5 & 76.5 & 74.1 & 52.6 & 0.73 & 0.44 & 0.64 & e\\
1188-0096357 & 05 46 33.09 +28 53 06.91   & 5 & 228.6 & 208.2 & 194.9 & 1.01 & 0.61 & 0.62 & c\\
1189-0098019 & 05 44 45.47 +28 55 59.58   & 5 & 25.7 & 65.3 & 49.1 & 0.60 & 0.67 & 0.89 & c/SDSS\\
1189-0098459 & 05 45 34.05 +28 59 58.33   & 5 & 243.0 & 245.6 & 269.2 & 1.21 & 0.48 & 0.70 & e/SDSS\\
1189-0098521 & 05 45 40.95 +28 55 29.41   & 5 & 195.6 & 176.2 & 187.8 & 0.45 & 0.45 & 0.67 & c\\
1189-0098637 & 05 45 56.05 +28 56 27.47   & 5 & 263.4 & 281.0 & 289.4 & 0.52 & 0.73 & 1.14 & c\\
1189-0098790 & 05 46 14.06 +28 54 22.38   & 5 & 195.5 & 229.6 & 170.5 & 0.37 & 0.35 & 0.49 & c\\
1190-0098093 & 05 45 21.49 +29 05 33.14   & 4 & 78.3 &  & 85.6 & 0.64 &  & 1.55 & c/SDSS\\
1190-0098147 & 05 45 26.81 +29 05 42.97   & 5 & 266.8 & 241.1 & 258.7 & 1.24 & 0.66 & 1.17 & e/SDSS\\
1190-0098184 & 05 45 30.60 +29 05 07.27   & 5 & 190.0 & 164.6 & 169.6 & 0.52 & 0.60 & 1.00 & e/SDSS\\
1191-0099312 & 05 45 51.34 +29 07 20.39   & 5 & 237.9 & 203.0 & 217.6 & 0.51 & 0.43 & 0.44 & e/SDSS\\
1251-0258748 & 17 33 39.50 +35 11 23.06   & 5 & 119.4 & 107.2 & 124.9 & 0.55 & 1.02 & 1.36 & e/SDSS\\
1251-0259277 & 17 35 16.84 +35 09 32.76   & 5 & 324.9 & 277.4 & -90.0 & 0.45 & 0.54 & 0.46 & c \\
1252-0258837 & 17 34 35.38 +35 17 25.89   & 5 & 308.1 & 264.3 & 274.6 & 0.47 & 0.40 & 0.50 & c/SDSS\\
1253-0259828 & 17 34 20.46 +35 19 41.06   & 5 & 244.7 & 256.5 & 254.2 & 0.40 & 0.77 & 1.10 & c/SDSS\\
1254-0260615 & 17 34 14.52 +35 27 45.51   & 5 & 296.6 & 274.9 & 266.8 & 0.49 & 0.94 & 0.89 & e/SDSS\\
1254-0261066 & 17 35 42.50 +35 26 46.43   & 5 & 61.7 & 74.5 & 87.7 & 0.44 & 0.49 & 0.49 & c/e\\
1254-0261346 & 17 36 31.45 +35 25 41.54   & 5 & 60.8 & 93.3 & 98.7 & 0.39 & 0.35 & 0.39 & c\\
1256-0261542 & 17 34 58.90 +35 40 57.35   & 5 & 310.0 & 291.3 & 293.9 & 0.90 & 0.44 & 0.77 & e\\
1269-0556127 & 22 06 50.33 +36 57 38.01   & 5 & 122.6 & 105.5 & 121.9 & 0.59 & 0.37 & 0.53 & c\\
1270-0576341 & 22 06 10.07 +37 00 21.47   & 4 & 156.2 & 157.8 &  & 0.47 & 0.48 &  & e	\\
1270-0576530 & 22 06 34.86 +37 00 09.77   & 4 & 79.4 & 80.4 &  & 1.30 & 0.72 &  & e\\
1271-0601107 & 22 06 32.34 +37 06 09.04   & 5 & 31.2 & 35.3 & 4.9 & 0.39 & 0.59 & 0.70 & c\\
1271-0601311 & 22 06 57.54 +37 10 04.65   & 5 & 268.9 & 272.7 & 277.3 & 0.51 & 0.85 & 1.34 & c\\
1271-0601577 & 22 07 25.96 +37 08 38.91   & 5 & 257.2 & 235.6 & 234.7 & 0.68 & 0.42 & 0.50 & e\\
1304-0186904 & 08 01 13.07 +40 24 24.51   & 5 & 282.5 & 250.8 & 230.5 & 0.37 & 0.46 & 0.44 & c/SDSS\\
\hline	
	\end{tabular}
\end{table*}

% Don't change these lines
\bsp	% typesetting comment
\label{lastpage*}
\end{document}